\documentclass{aa}  

\usepackage{graphicx}
\usepackage[dvipsnames]{xcolor}
\usepackage{caption}
\usepackage{multirow}
\usepackage{txfonts}
\usepackage[colorlinks,allcolors=blue]{hyperref}

\newcommand\figsize{0.45}
\newcommand{\rahour}{\hbox{\ensuremath{^{\rm h}}}}
\newcommand{\ramin}{\hbox{\ensuremath{^{\rm m}}}}
\newcommand{\xmm}{{\it XMM-Newton}\xspace}
\newcommand{\chandra}{{\it Chandra}\xspace}

\newcommand{\swift}{{\it Swift}\xspace}
\newcommand{\gaia}{{\it Gaia}\xspace}

\usepackage{threeparttable}

\usepackage{txfonts}
\begin{document}

   \title{Discovery of periodicities in two highly variable intermediate polars towards the Galactic Center}

    \author{Samaresh Mondal\inst{1},
          Gabriele Ponti\inst{1,2},
          Frank Haberl\inst{2},
          Kaya Mori\inst{3},
          Nanda Rea\inst{4,5}, 
          Mark R. Morris\inst{6},
          Sergio Campana\inst{1}, and
          Konstantina Anastasopoulou \inst{1,7}
          }

   \institute{$^1$INAF – Osservatorio Astronomico di Brera, Via E. Bianchi 46, 23807 Merate (LC), Italy
             \email{samaresh.mondal@inaf.it}\\
             $^2$Max-Planck-Institut für extraterrestrische Physik, Gießenbachstraße 1, 85748, Garching, Germany\\
             $^3$Columbia Astrophysics Laboratory, Columbia University, New York, NY 10027, USA\\
             $^4$Institute of Space Sciences (ICE, CSIC), Campus UAB, Carrer de Can Magrans s/n, E-08193 Barcelona, Spain\\
             $^5$Institut d’Estudis Espacials de Catalunya (IEEC), Carrer Gran Capità 2–4, E-08034 Barcelona, Spain\\
             $^6$Department of Physics and Astronomy, University of California, Los Angeles, CA 90095-1547, USA\\
             $^7$Harvard \& Smithsonian Center for Astrophysics, Cambridge, MA 20138, USA
             }

   \date{Received XXX; accepted YYY}
   \authorrunning{Mondal et al.}
   \titlerunning{Two variable IPs towards the GC}

 
  \abstract
   {}
   {We are performing a systematic analysis of X-ray point sources within 1\degr.5 of the Galactic center using archival \xmm data. While doing so, we discovered Fe $K_{\alpha}$ complex emission and pulsation in two highly variable sources (4XMM J174917.7--283329, 4XMM J174954.6--294336). In this work, we report the findings of the X-ray spectral and timing studies.}
   {We performed detailed spectral modeling of the sources and searched for pulsation in the light curves using Fourier timing analysis. We also searched for multi-wavelength counterparts for the characterization of the sources.}
   {The X-ray spectrum of 4XMM J174917.7--283329 shows the presence of complex Fe K emission in the 6--7 keV band. The equivalent widths of 6.4 and 6.7 keV lines are $99^{+84}_{-72}$ and $220^{+160}_{-140}$ eV, respectively. The continuum is fitted by a partially absorbed \texttt{apec} model with plasma temperature of $kT=13^{+10}_{-2}$ keV. The inferred mass of the white dwarf (WD) is $0.9^{+0.3}_{-0.2}\ M_{\odot}$. We detected pulsations with a period of $1212\pm3$ s and a pulsed fraction of $26\pm6\%$. 
   
   The light curves of 4XMM J174954.6--294336 display asymmetric eclipse and dipping behaviour. To date, this is only the second intermediate polar that shows a total eclipse in X-rays. The spectrum of the sources is characterized by a power-law model with photon index $\Gamma=0.4\pm0.2$. The equivalent widths of the fluorescent (6.4 keV) and Fe XXV (6.7 keV) iron lines are $171^{+99}_{-79}$ and $136^{+89}_{-81}$ eV, respectively. The continuum is described by emission from optically thin plasma with a temperature of $kT\sim35$ keV. The inferred mass of the WD is $1.1^{+0.2}_{-0.3}\ M_{\odot}$. We discovered coherent pulsations from the source with a period of $1002\pm2$ s. The pulsed fraction is $66\pm15\%$.}
   {The spectral modeling indicates the presence of intervening clouds with high absorbing column density in front of both sources. The detected periodic modulations in the light curves are likely to be associated with the spin period of WDs in magnetic cataclysmic variables. The measured spin period, hard photon index, and equivalent width of the fluorescent Fe $K_{\alpha}$ line are consistent with the values found in intermediate polars. While 4XMM J174954.6--294336 was already previously classified as an intermediate polar, we also suggest 4XMM J174917.7--283329 as a new intermediate polar. The X-ray eclipses in 4XMM J174954.6--294336 are most likely caused by a low-mass companion star obscuring the central X-ray source. The asymmetry in the eclipse is likely caused by a thick bulge that intercepts the line of sight during the ingress phase but not during the egress phase located behind the WD along the line of sight.}


   \maketitle
   
%
\section{Introduction}
Accreting white dwarf (WD) binaries are abundant in our universe \cite[see][for a recent review]{mukai2017}. WDs are a common endpoint of intermediate and low-mass stars, and many stars are born in a binary system with small separations that go through one or more mass transfer phases. Accreting WD binaries are categorized into two types, mainly on the basis of the companion star, which feeds the central X-ray source via Roche lobe overflow. Cataclysmic variables (CVs) have an early-type main-sequence donor, and symbiotic systems have a late-type giant donor. Understanding the long-term evolution of CVs is necessary for studying the progenitors of Type Ia supernovae and for future detection of gravitational wave sources by \emph{LISA} in the millihertz band \citep{melani2000,zuo2020}. Further, CVs are categorized into two types, non-magnetic and magnetic. Most of the hard X-ray emission from the Galactic center (GC) is expected to be produced by magnetic CVs \citep{revnivtsev2009,hong2009}. In magnetic CVs, the matter from the companion star is funneled through the magnetic field lines to the polar regions of the WD \citep{copper1990,petterson1994}. The in-falling material reaches a supersonic speed of 3000--10000 km s$^{-1}$, creating a shock front above the star and emitting thermal X-rays \citep{aizu1973}. There are two types of magnetic CVs: intermediate polars (IPs) and polars. IPs have a non-synchronous orbit with a WD surface magnetic field strength of $\sim$0.1--10 MG; they emit an ample amount of hard X-rays (20--40 keV). Polars are magnetically locked binary systems that have synchronized orbits with a strong magnetic field of 10--200 MG. Polars have softer X-ray spectra, $kT\sim5-10$ keV, due to faster cyclotron cooling \citep{mukai2017}.

A large number of CVs were detected through all-sky surveys such as performed by \emph{ROSAT} \citep{beuermann1999}, \emph{INTEGRAL} \citep{barlow2006} and \emph{Swift}-BAT \citep{baumgartner2013}. The 77-month \emph{Swift}-BAT catalogue, whose sky coverage is relatively uniform, lists around 81, of which roughly half are confirmed to be IPs \citep{baumgartner2013}. There are also deeper surveys focusing on a small part of the sky; for example, \citet{pretorius2007} exploited the \emph{ROSAT} all-sky survey, which was deeper near the north ecliptic pole, to infer the space density of CVs. Many star clusters are also prime targets for finding CVs. \citet{gosnell2012} discovered a candidate CV in the metal-rich open cluster NGC 6819 using \xmm. Globular clusters have been considered to host a large number of CVs; for example, among the X-ray sources in 47 Tuc \citep{grindlay2001a}, about 30 are considered likely CVs \citep{edmonds2003a,edmonds2003b} and in the Globular Cluster NGC 6397, nine likely to be CVs \citep{grindlay2001b}.

CVs have recently been discussed many times in the context of the GC \citep{krivonos2007,revnivtsev2009,hong2012,ponti2013,perez2015,hailey2016}. The diffuse hard X-ray emission in the GC and disk (the latter is termed as the Galactic ridge X-ray emission, or GRXE; \citealt{warwick1985}) is from a population of unresolved, faint point sources, including CVs \citep{revnivtsev2009,yamauchi2016}. However, the contribution from different types of sources and different types of CVs is still an open question. The only unambiguous way to constrain the CV population in the GC, ridge, and bulge is to analyze the individual X-ray point sources using spectra and light curves and identify them. Furthermore, estimating the X-ray-to-optical flux ratio by finding multi-wavelength counterparts can help to determine the source type. \citet{muno2003} detected 2350 X-ray point sources in the $17\arcmin\times17\arcmin$ field around Sgr A$^{*}$ and found that more than half of the sources are very hard, with photon index $\Gamma<1$, indicating magnetic CVs. \citet{yuasa2012} fitted the spectra of the Galactic ridge and bulge regions with a two-component spectral model and found the hard spectral component consistent with magnetic CVs of average mass $0.66^{+0.09}_{-0.07}\ M_{\odot}$. 

We are systematically studying X-ray point sources in the GC to understand the different types of X-ray binary populations. While doing this analysis, we found two relatively faint sources that display iron complex emission in X-ray spectra and periodicities in the light curves. In this paper, we report the X-ray spectral modeling, periodicities, and characterization of the two X-ray point sources in the GC. The coordinates of the sources are $(\alpha,\delta)_{\rm J200}$ = (17\rahour\,49\ramin\,17\fs7, --28\degr\,33\arcmin\,29\arcsec) and (17\rahour\,49\ramin\,54\fs6, --29\degr\,43\arcmin\,36\arcsec); both of these sources are listed in the 4XMM-DR11 catalogue as 4XMM J174917.7--283329 and 4XMM J174954.6--294336 \citep{webb2020}. 4XMM J174917.7--283329 is a newly identified point source with the detection of iron 6.4 and 6.7 keV lines and pulsations in the X-ray light curves. The source 4XMM J174954.6--294336 was first observed by \chandra during the Bulge Latitude Survey and then detected in Galactic Bulge Survey \citep{jonker2014}; later subsequently detected by \swift and \xmm. An association of a faint optical counterpart with an orbital period of 0.3587 days was identified by \citet{udalski2012}. A periodicity of 503.3 s was also detected in the optical light curve, which was interpreted as spin period \citep{johnson2017}. In this paper, we provide the actual spin period of the WD.

\section{Observations and data reduction}
This work is based on archival \xmm observations of the GC \citep{ponti2015,ponti2019}. The details of the observations are listed in Table \ref{table:obs_tab}. The observation data files were processed using the \xmm \citep{jansen2001} Science Analysis System (SASv19.0.0)\footnote{https://www.cosmos.esa.int/web/xmm-newton/sas}. We used the SAS task \texttt{barycen} to apply the barycentre correction to the event arrival times. We only selected events with PATTERN$\le4$ and PATTERN$\le12$ for EPIC-pn and EPIC-MOS1/MOS2 detectors, respectively. The source and background products were extracted from circular regions of 25\arcsec\ radius. The background products were extracted from a source-free area. The spectrum from each detector (pn, MOS1, MOS2) was grouped to have a minimum of 20 counts in each energy bin. The spectral fitting was performed in {\sc xspec} \citep{arnaud1996}, and we applied the $\chi^2$ statistic. The spectra from observations of EPIC-pn, MOS1, and MOS2 detectors were fitted simultaneously. While fitting the data simultaneously, we add a constant term for cross-calibration uncertainties, fixed to unity for EPIC-pn, and allowed to vary for MOS1 and MOS2. The best-fit parameters are listed in Table\,\ref{table:fit_tab} with the quoted errors at the 90\% significance level.

\begin{table}
\caption{The details of observations.}
\label{table:obs_tab}
\setlength{\tabcolsep}{1.5pt}                   
\renewcommand{\arraystretch}{1.5}               
\centering
\begin{tabular}{c c c c c}
\hline \hline
Name & ObsID & Date & Exposure & Total count\\ 
\hline
\multirow{3}{*}{J174917.7} & 0410580401 & 22-09-2006 & 31.6 ks & -/-/32 \\
& 0410580501 & 26-09-2006 & 31.1 ks & -/-/30\\
& 0801681301 & 07-10-2017 & 25.0 ks & 551/700/611 \\ 
\hline

\multirow{2}{*}{J174954.6} & 0801681401 & 07-10-2017 & 25.0 ks & 463/-/290 \\ 
& 0801683401 & 06-04-2018 & 26.0 ks & 800/316/317 \\ 
\hline

\end{tabular}
\tablefoot{The details of \xmm data analyzed in this paper. The columns represent  the name of the source, observation id, observation date, exposure time, and total counts for the pn/MOS1/MOS2 detectors.}
\end{table}

\section{Results}
\subsection{X-ray spectra}
We performed a detailed spectral analysis of the sources 4XMM J174917.7--283329 and 4XMM J174954.6--294336. We tested various phenomenological models to fit the spectra as well as a physical model to constrain the mass of the central WD. The results from the spectral fitting are described in the following subsections. All the spectral fitting models are convolved with a Galactic absorption component \texttt{tbabs} with the photoionization cross sections and abundance values from \cite{wilms2000}. 

\subsubsection{4XMM J174917.7--283329}
The source was observed three times by \xmm. The observations done on 22-09-2006 (ObsID: 0410580401) and 26-09-2006 (ObsID: 0410580501) were in timing mode and pointed at IGR J17497--2821, so the source was outside the field of view of the EPIC-pn and MOS1 detectors. In the case of the MOS2 detector, the source was marginally detected due to the high background and low flux state of the source. Hence we used the ObsIDs 0410580401 and 0410580501 to estimate the flux of the source only. Later the same field was observed by \xmm on 07-10-2017 (ObsID: 0801681301), in which the source was brighter and clearly detected by all three detectors. We used spectra from this observation for our detailed spectral modeling. 

\begin{figure}[]
\centering
\includegraphics[width=\figsize\textwidth]{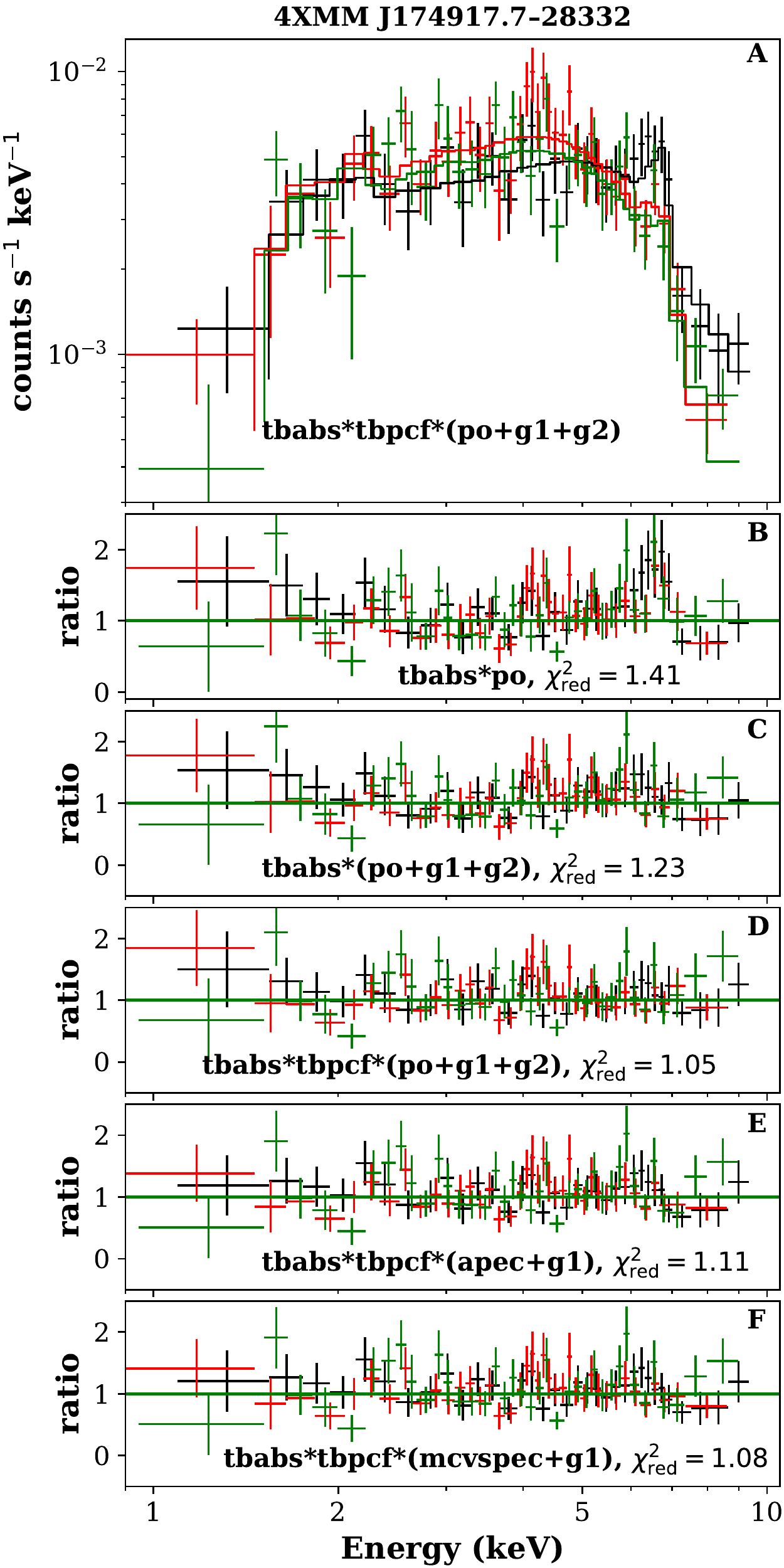}
\caption{The various spectral model fits to the spectra of 4XMM J174917.7--283329. The black, red, and green colors represent the spectra from EPIC-PN, MOS1, and MOS2 detectors, respectively. Panel A represents the best-fit spectral model overlaid on the data points. The lower panels indicate the ratio plot obtained from the fitting of various models. The various model components are, \texttt{tbabs}: Galactic absorption, \texttt{tbpcf}: absorption from medium partially covering the X-ray source, \texttt{po}: power-law continuum, \texttt{apec}: emission from collisionally-ionized diffuse gas, \texttt{mcvspec}: continuum emission from WD accretion column, \texttt{g1}: Gaussian line at 6.4 keV and \texttt{g2}: Gaussian line at 6.7 keV.}
\label{fig:src1_ratio}
\end{figure}

First, we fit the spectra with a simple absorbed power-law model. Fitting with this model indicates the source has a hard photon index with $\Gamma=0.9\pm0.2$ and shows the presence of excess emission in the 6--7 keV band, which is shown in panel B of Fig. \ref{fig:src1_ratio}. The resultant fit statistics is $\chi^2=152$ for 108 degrees of freedom (d.o.f.). The excess between 6 and 7 keV is fitted by adding two Gaussian lines at 6.4 keV ($\chi^2=146$ for 107 d.o.f. with 96.16\% detection significance in an F-test) and 6.7 keV ($\chi^2=136$ for 107 d.o.f. with 99.94\% detection significance in an F-test). We did not find any improvement in the fit after adding another Gaussian at 6.9 keV for the Fe XXVI line. The improvement in the fit after adding the lines is shown in panel C of Fig. \ref{fig:src1_ratio}. We left the width of the lines free but found them to be consistent with being narrow; therefore, we froze the width of the Gaussian lines to zero. While adding the two Gaussians, the statics of the spectral fit is significantly improved by $\Delta\chi^2=22$ for two additional d.o.f. The equivalent width and its 90\% error on the lines at 6.4 keV and 6.7 keV are $99^{+84}_{-72}$ eV, and $220^{+160}_{-140}$ eV, respectively. Next, we add a partial covering to the model, which represents the emission partially covered by the intervening medium in front of the source. The column density of the intervening medium is almost 5-9 times higher than the Galactic absorption. The Galactic absorption column density from the spectral fit is $N_{\rm H}\sim(3\pm0.7)\times10^{22}\ \rm cm^{-2}$. Adding the partial covering further improves the fit with $\Delta\chi^2=21$ for two additional d.o.f. The resultant fit is shown in panels A and D of Fig. \ref{fig:src1_ratio}. To estimate the temperature of the X-ray emitting plasma, we fit the spectra with the \texttt{apec} model together with the partial covering absorption. The \texttt{apec} uses both the shape of the continuum and the line ratio of 6.7 keV and 6.9 keV to estimate the plasma temperature. Furthermore, the \texttt{apec} model represents the emission from the ionized material. Therefore, it does not include the neutral iron $K_{\alpha}$ line emission at 6.4 keV. Hence we add a Gaussian line at 6.4 keV to the \texttt{apec} model. Fitting the spectrum with this model provides a best-fit plasma temperature of $kT=13^{+10}_{-2}$ keV. 

Next, we fit the data with a physically motivated model called \texttt{mcvspec}. The model is an evolution of the model presented in \citet{saxton2005} by Mori et al. (in preparation) and is available in {\sc xspec}. This model represents the emission from the surface of a WD. It only includes lines produced collisionally in an ionized, diffuse gas in the accretion column of the WD. Therefore, we again add a Gaussian at 6.4 keV for the neutral iron $K_{\alpha}$ line to take into account the X-ray reflection of the WD surface or pre-shock region. While doing the fit with this model, we freeze the magnetic field $B$ and the mass accretion flux $\dot m$ to values of 10 MG and 5 g cm$^{-2}$ s$^{-1}$, respectively, which are the values typically found in IPs. The WD mass obtained by fitting this model is $0.9^{+0.3}_{-0.2}\ M_{\odot}$.

\subsubsection{4XMM J174954.6--294336}
The field around 4XMM J174954.6--294336 was observed twice by \xmm. The observation done on 07-10-2017 (ObsID: 0801681401) was performed in full frame mode; however, the source fell into the chip gap of the MOS1 detector; therefore, we report only the analysis of the EPIC-pn and MOS2 detectors. The source was also observed by \xmm on 06-04-2018 (ObsID: 0801683401) by all three detectors. We noticed that between the 2017 and 2018 observations, the source flux varied by a factor of 1.45. However, the shapes of the continua are very similar. Therefore, we fit the combined spectra of 2017 and 2018 observations to gain statistics.

\begin{figure}[]
\centering
\includegraphics[width=\figsize\textwidth]{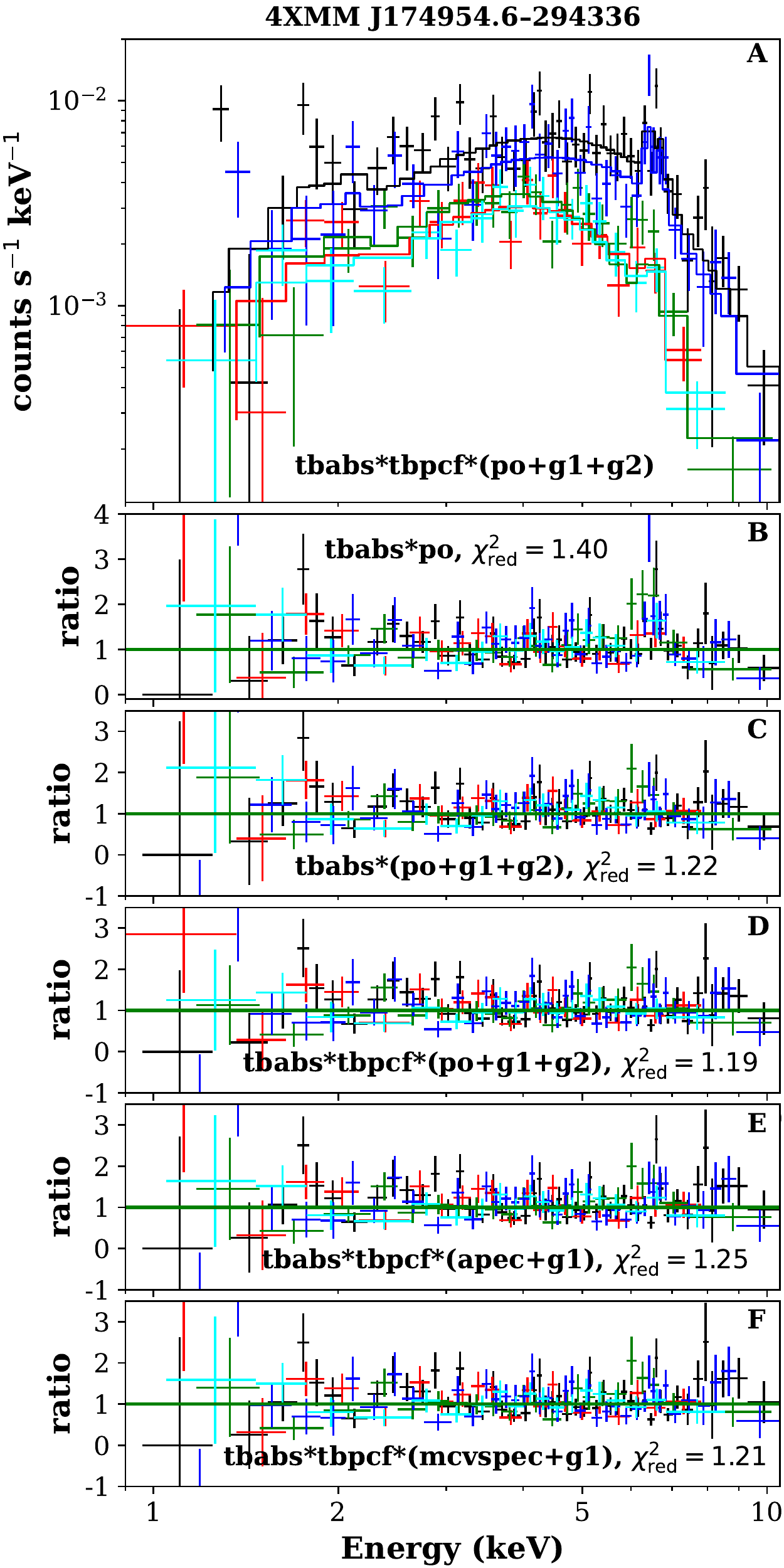}
\caption{Same as Fig. \ref{fig:src1_ratio} but for the source 4XMM J174954.6--294336. The black, red, and green data points are from the EPIC-pn, MOS1, and MOS2 detectors of ObsID 0801683401. The blue and cyan data points are from the EPIC-pn and MOS2 detectors of ObsID 0801681401.}
\label{fig:src2_ratio}
\end{figure}

Fitting an absorbed power-law model provides a best-fit photon index of $\Gamma=0.4\pm0.2$. Residuals around the iron line complex are clearly visible in the ratio plot, which is shown in panel B of Fig. \ref{fig:src2_ratio}. To resolve the excess in the 6--7 keV band, we add two Gaussians at 6.4 keV and 6.7 keV to the power-law model, which improves the fit by $\Delta\chi^2=29$ for two additional d.o.f. We performed an F-test, which gives a detection significance of 99.98\% and 99.86\% for the 6.4 keV and 6.7 keV lines, respectively. Further, adding another Gaussian at 6.9 keV for the Fe XXVI line does not improve the fit. The equivalent width of the lines at 6.4 keV and 6.7 keV are $171^{+99}_{-79}$ eV and $136^{+89}_{-81}$ eV, respectively. Next, we add a partial covering absorption model to the power-law continuum, which improves the fit marginally by $\Delta\chi^2=7$ for two additional d.o.f. However, we noticed while fitting with the \texttt{apec} and \texttt{mcvspec} continuum models that adding a partial covering absorption improves the fit by $\Delta\chi^2=42$ and 44, respectively, for two extra additional d.o.f. Fitting with the \texttt{apec} model provides a best-fit plasma temperature of $kT=35\pm17$ keV. Furthermore, we fit the spectra with the \texttt{mcvspec} model. Such as done for 4XMM J174917.7--283329 while fitting with the \texttt{mcvspec} model we freeze $B$ to 10 MG and $\dot m$ to 5 g cm$^{-2}$ and s$^{-1}$. The mass of the central compact object estimated from the \texttt{mcvspec} model is $1.1^{+0.2}_{-0.3}\ M_{\odot}$. 

\begin{table*}
\caption{The best-fit parameters of the fitted models.}
\label{table:fit_tab}
\setlength{\tabcolsep}{5pt}                   
\renewcommand{\arraystretch}{2.0}               
\centering
\begin{tabular}{c c c c c c c c c}
\hline\hline
 \multirow{2}{*}{\texttt{tbabs}*Model} & $N_{\rm H}$ & $N_{\rm H,pcf}$ & $pcf$ & $\Gamma/kT/M$ & $N_1$ & $N_{\rm g1}$ & $N_{\rm g2}$ & $\chi^2/dof$\\ 
 & $\times10^{22}$ & $\times10^{22}$ &&&& $\times10^{-6}$ & $\times10^{-6}$ &\\ \hline

 \multicolumn{9}{c}{4XMM J174917.7--283329}\\ \hline
 \texttt{po} & $3.1^{+0.6}_{-0.5}$ &&& $0.9\pm0.2$ & $9^{+4}_{-2}\times10^{-5}$ &&& 152/108\\ \hline
 \texttt{po+g1+g2} & $3.3^{+0.6}_{-0.5}$ &&& $1.0\pm0.2$ & $1.0^{+0.4}_{-0.3}\times10^{-4}$ & $2\pm1$ & $4\pm2$ & 130/106\\ \hline
 \texttt{tbpcf*(po+g1+g2)} & $4\pm1$ & $27^{+17}_{-14}$ & $0.8^{+0.1}_{-0.2}$ & $2.3\pm0.6$ & $2^{+6}_{-1}\times10^{-3}$ & $2^{+2}_{-1}$ & $4\pm3$ & 109/104\\ \hline
 \texttt{tbpcf*(apec+g1)} & $2.9\pm0.7$ & $15^{+11}_{-6}$ & $0.6\pm0.1$ & $13^{+10}_{-2}$ & $1.2\pm0.2\times10^{-3}$ & $2\pm1$ && 117/105\\ \hline
 \texttt{tbpcf(mcvspec+g1)} & $3.0^{+0.6}_{-0.7}$ & $16^{+11}_{-6}$ & $0.6\pm0.1$ & $0.9^{+0.3}_{-0.2}$ & $1.3^{+0.5}_{-0.4}\times10^4$ & $2\pm1$ && 113/105\\ \hline

 \multicolumn{9}{c}{4XMM J174954.6--294336}\\ \hline
 \texttt{po} & $2.8^{+0.7}_{-0.6}$ &&& $0.4\pm0.2$ & $4^{+2}_{-1}\times10^{-5}$ &&& 209/149 \\ \hline
 \texttt{po+g1+g2} & $3.0^{+0.8}_{-0.6}$ &&& $0.6\pm0.2$ & $4\pm1\times10^{-5}$ & $3\pm1$ & $3\pm2$ & 180/147\\ \hline
 \texttt{tbpcf*(po+g1+g2)} & $2^{+1}_{-2}$ & $9^{+13}_{-6}$ & $0.6\pm0.3$ & $1.1^{+0.6}_{-0.4}$ & $1.2^{+2.7}_{-0.7}\times10^{-4}$ & $4\pm2$ & $3\pm2$ & 173/145\\ \hline
 \texttt{tbpcf*(apec+g1)} & $3.0^{+0.9}_{-1.0}$ & $16^{+12}_{-6}$ & $0.7\pm0.1$ & $35\pm16$ & $1.2^{+0.3}_{-0.2}\times10^{-3}$ & $4\pm2$ && 182/146\\ \hline
 \texttt{tbpcf(mcvspec+g1)} & $3.0^{+0.9}_{-1.0}$ & $15^{+11}_{-5}$ & $0.7\pm0.1$ & $1.1^{+0.2}_{-0.3}$ & $9^{+4}_{-5}\times10^3$ & $4\pm2$ && 177/146\\ \hline
 
\hline
\end{tabular}
\tablefoot{$N_{\rm H}$ is given in units of $10^{22}$ cm$^{-2}$, $kT$ in keV and $M$ in $M_{\odot}$. For the \texttt{apec} and \texttt{mcvspec} models, the metal abundance value is frozen to 1.0. In the \texttt{mcvspec} model, we freeze $B$ and $\dot m$ to 10 MG and 5 gm cm$^{-2}$ s$^{-1}$, typical values for low magnetized WDs. Due to a lack of good-quality data, we had to freeze the centroid of the Gaussian lines; otherwise, it takes random values while fitting.}
\end{table*}

\subsection{Periodicity search}
We computed the power spectral densities (PSD) to search for periodicities in the 1--10 keV light curves. For our PSD analysis, we used EPIC-pn light curves only, as it has the shortest frame time that allows us to probe a higher frequency range. Next, to refine the detected period and estimate the error, we search for maximum $\chi^2$ as a function of the period using the FTOOL \texttt{efseach}. Then we used the refined period to fold the light curve and estimate the pulsed fraction in the 1--10 keV band. The pulsed fraction was estimated by using the formula $\rm PF=\frac{F_{max}-F_{min}}{F_{max}+F_{min}}\times100\%$, where $\rm F_{\rm max}$ and $\rm F_{\rm min}$ are the maximum and minimum of the normalized intensity, respectively.

\subsubsection{4XMM J174917.7--283329}
The left top, middle, and bottom panels of Fig. \ref{fig:psd} show the PSD, $\chi^2$ search, and the folded light curve of source 4XMM J174917.7--283329, respectively. The PSD shows a peak at frequency $8.39\times10^{-4}$ Hz. We used this frequency as an input in the \texttt{efsearch} algorithm. The refined period and its 90\% ($\Delta\chi^2=\pm2.7$) error is $1212\pm3$ s. Further, we folded the light curve with the given period, and the estimated pulsed fraction is $26\pm6$\%.

\subsubsection{4XMM J174954.6--294336}
\begin{figure}[]
\centering
\includegraphics[width=\figsize\textwidth]{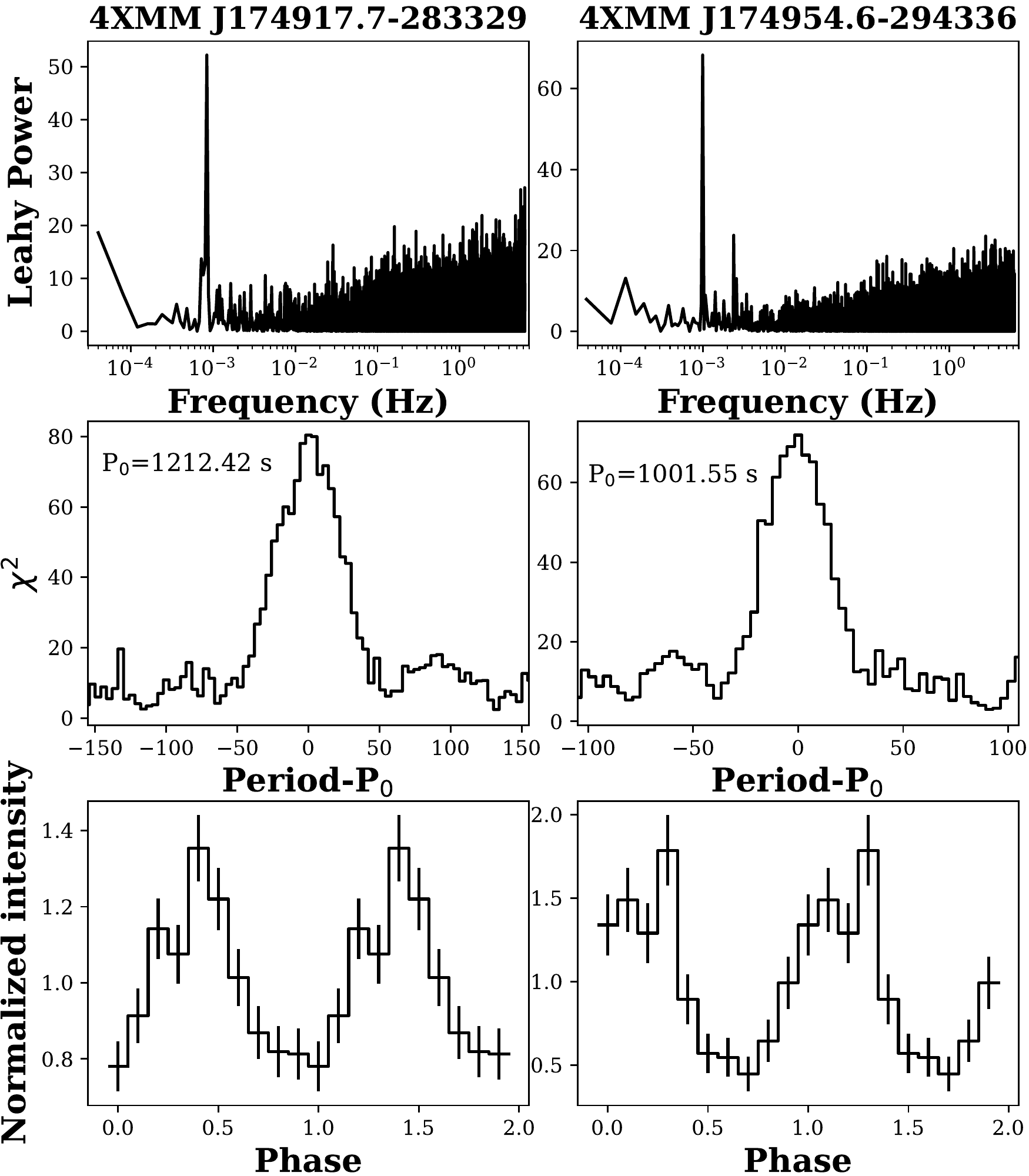}
\caption{The top panels shows the periodogram in Leahy normalization obtained from the EPIC-pn light curve of the source 4XMM J174917.7--283329 (left panel) and 4XMM J174954.6--294336 (right panel). The middle panels show the $\chi^2$ analysis using the FTOOL \texttt{efsearch}. The bottom panels show the folded light curves.}
\label{fig:psd}
\end{figure}

\begin{figure}[]
\centering
\includegraphics[width=\figsize\textwidth]{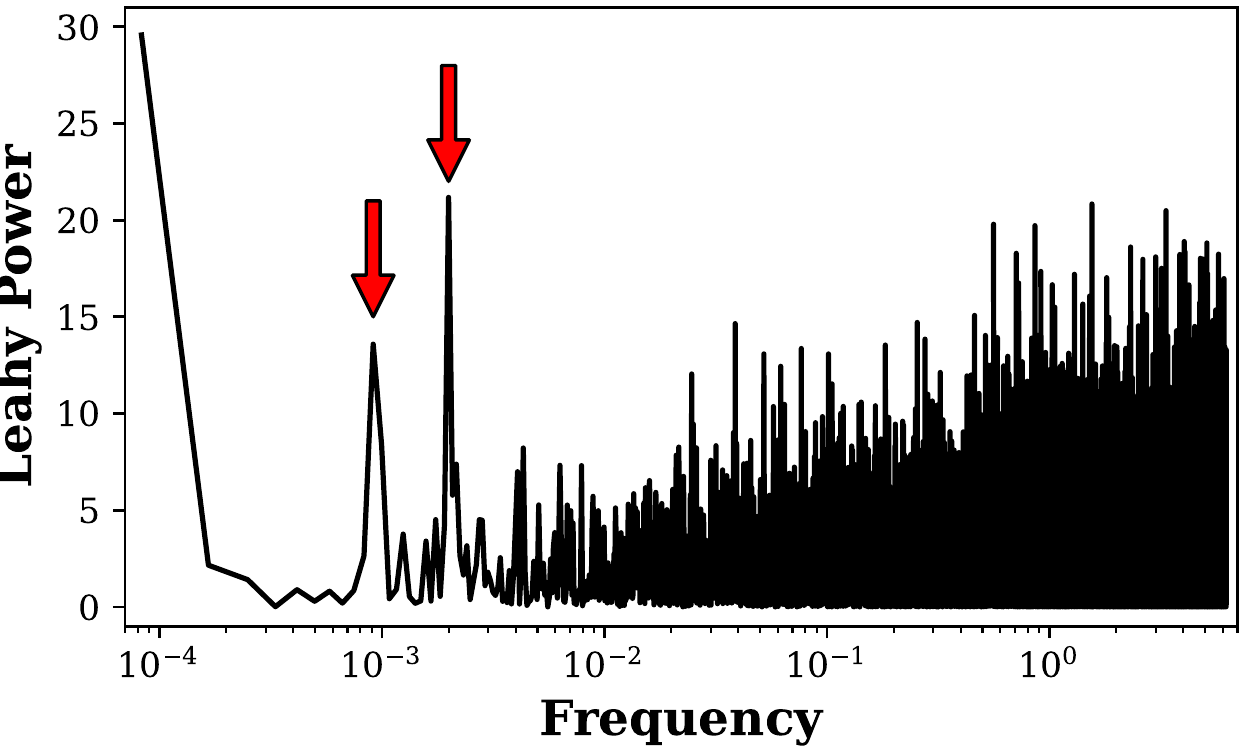}
\caption{The periodogram of 4XMM J174954.6--294336, obtained using the initial 10 ks of the observation with ObsID 0801681401. Two peaks were observed but not at a very high significance level.}
\label{fig:psd1}
\end{figure}

The right panels of Fig. \ref{fig:psd} show the results obtained from the timing analysis of 4XMM J174954.6--294336 using ObsID 0801683401. The PSD shows a peak at $9.98\times10^{-4}$ Hz. The estimated period and error from the \texttt{efsearch} analysis is $1002\pm2$ s. The pulsed fraction of the source is $66\pm15$\%. We noticed that the eclipse duration of 2500 s at the end of the light curve introduces a spurious signal in the PSD at a frequency of $2.38\times10^{-3}$ Hz. Further, we analyzed the light curve from the ObsID 0801681401 and did not find any clear signal in the PSD at the corresponding frequency of 1002 s period. In this observation, we noticed that the source light curve shows dipping behaviour caused by absorption. Therefore, the pulsed signal is likely lost due to the variation introduced by the absorption. In fact, by computing the FFT using the initial 10 ks of this light curve which is unaffected by the absorption, the PSD shows two peaks at frequencies $9.13\times10^{-4}$ Hz, which is consistent with the 1002 s period and $1.99\times10^{-3}$ Hz which is likely the first harmonic of the fundamental period (Fig. \ref{fig:psd1}).

\subsection{The long-term X-ray variability}
We constructed the long-term light curve spanning over the time scale of ten years by searching for counterparts in \swift 2SXPS\footnote{https://www.swift.ac.uk/2SXPS/} \citep{evans2020} and \chandra CSC 2.0\footnote{https://cxc.cfa.harvard.edu/csc2/} catalogues \citep{evans2010}. 
\subsubsection{4XMM J174917.7--283329}
Figure \ref{fig:long_flux} shows the long-term flux variation of the source 4XMM J174917.7--283329 (top panel). The source has been detected multiple times by \xmm and \swift and displays a flux variation by a factor of six or more over the timescale of ten years. 

\subsubsection{4XMM J174954.6--294336}
\begin{figure}[]
\centering
\includegraphics[width=\figsize\textwidth]{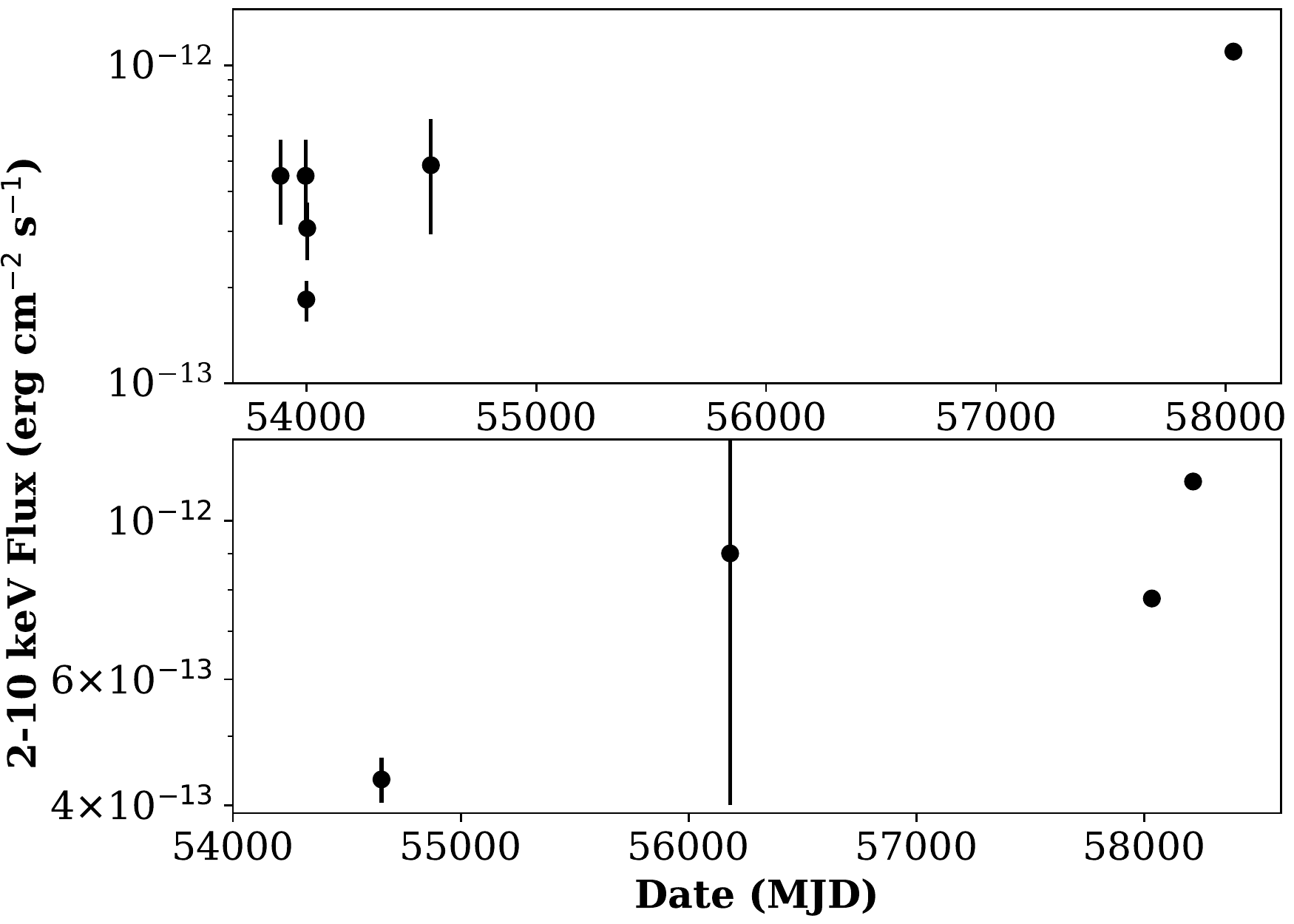}
\caption{The long-term flux variation of 4XMM J174917.7--283329 (top panel) and 4XMM J174954.6--294336 (bottom panel). Both sources show significant flux variability.}
\label{fig:long_flux}
\end{figure}

\begin{figure*}[]
\centering
\includegraphics[width=0.95\textwidth]{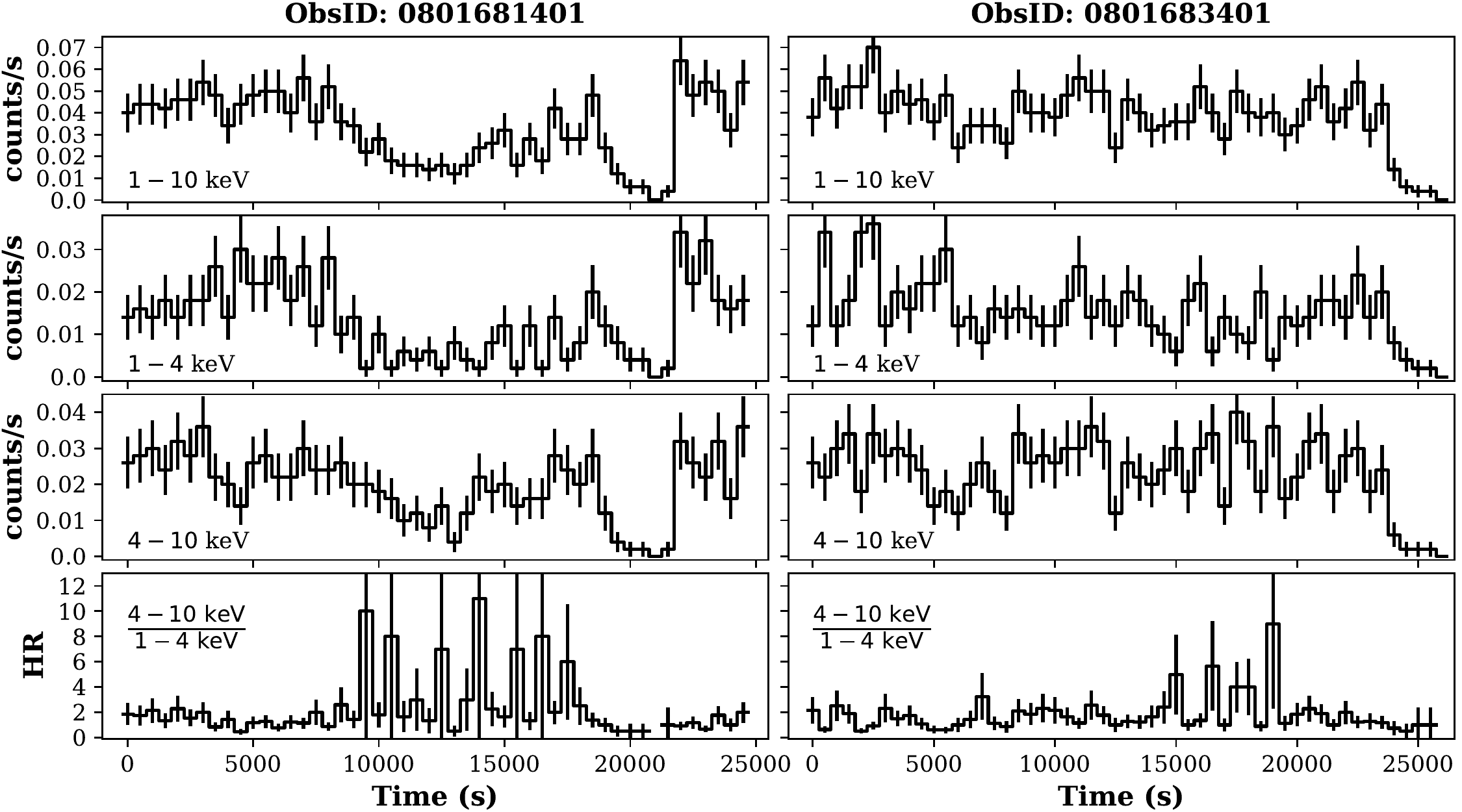}
\caption{The EPIC-pn light curve of the source 4XMM J174954.6--294336 with 500s time bin in various energy bands 1--10 keV (top panels), 1--4 keV, 4--10 keV (middle panels), and hardness ratio plot in bottom panels.}
\label{fig:src2_lc}
\end{figure*}

Figure \ref{fig:long_flux} bottom panel shows the long-term light curve of the source 4XMM J174954.6--294336. The flux of the source varies by a factor of three. Figure  \ref{fig:src2_lc} shows the EPIC-pn 1--10 keV, 1--4 keV, and 4--10 keV light curves. The light curves were binned with a time resolution of 500 s. The light curves show remarkable features with a long-term variation and two eclipses near the end of the observations. Obscuration of the central X-ray source by the companion star likely causes the eclipses. In the first observation, the 1--4 keV band (middle panel of Fig. \ref{fig:src2_lc}) light curve also shows very short-term variation associated with the absorption due to dipping behaviour before entering into the eclipses. During the dipping activity, the soft X-ray photons (1--4 keV) are absorbed more than the hard 4--10 keV photon leading to an increase in hardness ratio (bottom panel of Fig. \ref{fig:src2_lc}); this indicates an absorption related origin.

\section{Discussion}
\subsection{4XMM J174917.7--283329}
The hard X-ray spectrum of 4XMM J174917.7--283329 can be characterized by a power law with photon index $\Gamma=0.9\pm0.2$. The presence of excess emission in the 6--7 keV band can be attributed to the iron $K_{\alpha}$ complex. The equivalent width of the 6.4 keV and 6.7 keV lines are $99^{+84}_{-72}$ eV and $220^{+160}_{-140}$ eV, respectively. Our spectral fitting indicates the presence of an absorbing medium close to the source with $N_{\rm H,pcf}\sim(1.5-3)\times10^{23}\ \rm cm^{-2}$ which partially absorbs the incoming X-ray photons. The plasma temperature of the accreting material is $kT=13^{+10}_{-2}$ keV. The central WD mass estimated from fitting a physical model is $0.9^{+0.3}_{-0.2}\ M_{\odot}$. The Galactic neutral atomic hydrogen column density towards the source is $1.1\times10^{22}\ \rm cm^{-2}$ ($N_{\rm H}=N_{\rm HI}+N_{\rm H2}$; \citealt{willingale2013}), which is lower than the absorption column density obtained from the X-ray spectral fitting. For the first time, we detected the spin period of the WD is $1212\pm3$ s. 

For better positional accuracy, we searched for an X-ray counterpart in the \chandra source catalogue; however, no \emph{Chandra} observation of this region has been performed so far. Two possible \gaia counterparts were found within 0.05\arcmin\ from the \xmm position with $G_{\rm mag}$ of 20.48 and 18.96. Both \gaia sources have a similar parallax of 0.48 mas, which translates to a distance of 2.08 kpc. The X-ray source flux varied by a factor of six over a time scale of ten years. The 2--10 keV luminosity variation of the source is (1--6)$\times10^{32}\rm \ erg\ s^{-1}$. 

\subsection{4XMM J174954.6--294336}
The spectra of 4XMM J174954.6--294336 are characterized by a hard power-law with a photon index of $\Gamma=0.4\pm0.2$, which is typically seen from accreting WDs. Moreover, a partially absorbed optically thin plasma of temperature $kT=35\pm16$ keV provides an adequate fit to the spectra. In addition to that, the spectra display the presence of fluorescent 6.4 keV and ionized 6.7 keV lines. The equivalent widths of the lines are $171^{+99}_{-79}$ eV and $136^{+89}_{-81}$ eV for 6.4 keV and 6.7 keV, respectively. The 6.4 keV line originates from the reflection from the surface of the WD or from the pre-shock region in the accretion column and typically has an equivalent width of 150 eV \citep{ezuka1999}.

The X-ray light curve shows coherent pulsations with a period of $1002\pm2$ s. The pulsation signal was suppressed in an earlier \xmm observation with ObsID 0801681401. This is due to the energy-dependent absorption dips (prominent in the 1--4 keV band, middle panel of Fig. \ref{fig:src2_lc}), which dilutes the coherent pulsations. However, the pulsations were marginally detected in the initial one-third of that observation, which is unaffected by the dips. These dips are believed to be caused by photoelectric absorption by surrounding material. In the later observation of 0801683401, the dipping phenomenon is not present in the soft band 1--4 keV light curve before the source goes into the eclipse phase. This suggests the dips are highly irregular and variable from orbit to orbit. Similar dipping behaviour was also detected in other X-ray eclipsing sources such as dwarf novae \citep{mukai2009} and low-mass X-ray binaries \citep{diaz2006,ponti2016}. It is well established that the dipping phenomena are seen in high inclination systems. The physical model for explaining this dipping behaviour is linked to the obscuration of the central X-ray source by absorbing material in the region where the stream of material from the companion star hits the outer rim of the accretion disk. This leads to the thickening of the disk rim with azimuth, generating a thick bulge where the stream hits the disk edge, and the dipping occurs when the bulge intercepts the line of sight to the central X-ray source \citep{white1985}. On the other hand, there is another physical picture of the dipping phenomena where the disk structure is fixed. The dipping activity originates from the interaction of matter that has been left of the stream above and below the accretion disk \citep{frank1987}.

The source 4XMM J174954.6--294336 is classified as a nova-like variable \citep{ritter2003}. The source was detected by \emph{Chandra} in the Galactic Bulge Survey (GBS) and is designated as CXOGBS J174954.5--294335 \citep{jonker2011,jonker2014}. \citet{udalski2012} did a systematic search for optical counterparts of GBS sources, and the source appeared in OGLE-IV fields. Two possible optical counterparts within 3\farcs9 from the \chandra source position were found: a variable red giant with a period of 31.65 days with $I_{\rm mag}=15.67$ and a fainter eclipsing binary with a period of 0.3587 days with $I_{\rm mag}=17.98$. The \chandra and \xmm source locations are consistent at 2$\sigma$ position uncertainty. The eclipsing binary system has observed a $V-I$ color magnitude of 1.52. \citet{britt2014} also did an optical search for the GBS sources using the Blanco 4 m Telescope at CTIO. An optical counterpart of $r_{\rm mag}=19.21$ was found associated with the X-ray source. Their optical light curve shows aperiodic variability of 0.4 mag and an eclipse of almost one magnitude depth. Given the detection of the eclipses in the X-ray light curves, it is very likely that the eclipsing binary with the period of $P_{\rm orb}=0.3587$ days is the actual optical counterpart of the X-ray source 4XMM J174954.6–294336. \citet{johnson2017} analyzed optical photometry data from DECam and OGLE. They obtained a similar orbital period as \citet{udalski2012} and discovered a spin period of the WD of 503.3 s. The detected 1002 s X-ray periodicity is consistent with twice the optical period of 503.3 s. In a later \xmm observation (ObsID: 0801681401), the peaks close to 1002 and 503 s were detected in the initial 10 ks observation (Fig. \ref{fig:psd1}). This indicates that the true spin period is 1002 s. Furthermore, \citet{johnson2017} analyzed the data from \emph{Chandra} and detected a total X-ray eclipse. However, the \emph{Chandra} data do not have enough signal-to-noise to detect the asymmetric shape of the eclipses and the iron line complex. We searched for counterparts in the \gaia catalogue \citep{gaia2016,gaia2022}. An optical source with \gaia $G_{\rm mag}=18.97$ is consistent in position with the eclipsing system. The estimated parallax obtained from the \emph{Gaia} data is $0.61\pm0.34$ mas which translates into the distance to the source of $\sim1.64^{+2.06}_{-0.59}$ kpc.

In each of the two \xmm observations, we detected an X-ray eclipse in which the count rate went to zero. In one observation, we detected a total X-ray eclipse; however, in the later observation, only the ingress phase was caught. So far, only a few accreting WDs are known to display eclipses in X-rays \citep{hellier1997,schwope2001,pandel2002,ramsay2007,mukai2009} and 4XMM J174954.6--294336 is only the second IP after XY Ari \citep{hellier1997} that shows complete eclipses in X-rays. X-ray eclipses are a powerful diagnostic tool to constrain the geometry of the binary system. The duration of the eclipse ingress (the time interval between first and second contact) and egress (the time interval between third and fourth contact) is used to estimate the fractional area $f$ of the X-ray emitting region on the WD surface. So far, $f$ was constrained only for one IP \citep{hellier1997}. Typically the ingress and egress times are of a few seconds. The ingress phase of 4XMM J174954.6--294336 lasted around 1500 s, which is much larger than previously found in eclipsing WDs. We detected one complete eclipse, which is asymmetric. The egress phase takes less than 500 s. Further, it is noticeable that the asymmetry is more pronounced in the soft 1--4 keV band than in the hard 4--10 keV band, suggesting an absorption-related origin. A similar asymmetric eclipse behaviour was seen in the eclipsing polar HU Aqr \citep{schwope2001} in which the ingress took longer because of the effect of absorption dips which is discussed previously. At the same time, the egress is clean and lasts only 1.3 s. Asymmetric eclipses are more common in eclipsing high-mass X-ray binaries such as 4U 1700--37 \citep{haberl1989} and Vela X-1 \citep{haberl1990,falanga2015}. \citet{falanga2015} studied a sample of bright high-mass X-ray binaries using data from \emph{INTEGRAL} and found that the asymmetric shape is seen more clearly in the soft (1.3--3, 3--5, 5--12 keV) bands than in the hard (40--150 keV) band. They suggest that the asymmetry is caused by an increase in local absorption column density due to accretion wakes \citep{blondin1990,manousakis2012}. During the egress phase, the wake is located behind the compact object along the line of sight, thus not leading to any apparent increase in the local absorption column density. Therefore the egress phase is clean and much shorter than the ingress phase. However, the companion of 4XMM J174954.6--294336 is unlikely to be a high mass system as \citet{johnson2017} estimated the spectral type of the donor to be G3V–G5V from density-period relation, which is associated with a main sequence star of $0.9-1.0\ M_{\odot}$.

The estimated distance of 1.64 kpc to the source suggests a  2--10 keV luminosity of (1--4)$\times10^{32}\ \rm erg\ s^{-1}$. On the other hand, the mean Galactic absorption column density towards the source location is $1.0\times10^{22}\ \rm cm^{-2}$ \citep{willingale2013}, which is lower but within a factor of two of the value obtained from the spectral fitting of the source.

Optical measurements of a sample of 32 sources indicate the mean WD mass among CVs is $0.83\pm0.23\ M_{\odot}$ \citep{zorotovic2011}. On the other hand, using \emph{RXTE} observations of 20 magnetic CVs, \citet{ramsay2000} derived a mean mass of $0.85\pm0.21\ M_{\odot}$ and $0.80\pm0.14\ M_{\odot}$, for IPs and polars, respectively. In recent years \emph{NuSTAR} observations have been effective in measuring mass due to the high energy coverage and sensitivity of the instrument \citep{hailey2016,suleimanov2016,suleimanov2019,shaw2020}. \citet{shaw2020} measured the mass of 19 IPs using \emph{NuSTAR} and found the mean mass to be $0.77\pm0.10\ M_{\odot}$. These studies suggest that CVs, IPs, and polars have similar masses but higher than the pre-CVs and isolated WDs, giving rise to a WD mass problem. The pre-CV population have mean mass of $0.67\pm0.21\ M_{\odot}$ \citep{zorotovic2011} and isolated WDs have a mean mass of $0.53\pm0.15\ M_{\odot}$ \citep{kepler2016}. Understanding the mass distribution of accreting WDs is crucial in explaining the formation and evolution of magnetic and non-magnetic CVs. We obtained the mass of the WDs by fitting a physical spectral model to the spectra. For both sources, the estimated mass is consistent with the mean mass of CVs. While doing the spectral fit, we freeze the $B$ and $\dot m$ due to degeneracy; this may have some effect on the estimation of the mass. Few IPs in the GC and bulge regions are found to have masses above $1\ M_{\odot}$ such as IGR J1807--4146 ($1.06^{+0.19}_{-0.10}\ M_{\odot}$; \citealt{coughenour2022}), 4XMM J174033.8–301501 ($1.05^{+0.16}_{-0.21}\ M_{\odot}$; \citealt{mondal2022}), and CXO J174517.0--321356.5 ($1.1\pm0.1\ M_{\odot}$; Vermette et al. in preparation).

\section{Conclusions}
In this paper, we performed detailed spectral and timing studies of two highly variable X-ray sources located within 1\degr.5 of the Galactic center. Furthermore, we characterize the sources using their multi-wavelength counterparts. 

The 1--10 keV spectra of 4XMM J174917.7–283329 can be characterized as emission from optically thin plasma with temperature $kT=13^{+10}_{-2}$ keV. In addition to that, a partial covering absorption with column density much higher than the Galactic value is required to fit the spectrum. The partial covering can be inferred as absorption due to circumstellar gas located close to the source. We estimate the mass of the central WD as $0.9^{+0.3}_{-0.2}\ M_{\odot}$. Our timing analysis revealed pulsations with a period of $1212\pm3$ s, and the long-term flux measurements suggest the source is highly variable.

The hard X-ray spectrum of 4XMM J174954.6–294336 resembles in shape the typical spectra seen from accreting WDs. The source was already identified as an IP with an orbital period of 0.3587 days. The X-ray spectra are well fitted by a model of optically thin plasma of $kT\sim35$ keV. The estimated mass of the WD is $1.1^{+0.2}_{-0.3}\ M_{\odot}$. We performed Fourier timing analysis and detected pulsations with a period of $1002\pm2$ s. The long-term observations indicate a flux variability by a factor of three. Since these types of sources are naturally variable, a flux variation of this amplitude is expected. In addition to that, the short-term X-ray light curves display complete eclipses and absorption dips. Due to the limited statistical quality of the data and the number of eclipses detected, a detailed phase-dependent study is not possible. Follow-up X-ray observations of 4XMM J174954.6–294336 with more eclipses detected will help to constrain the binary system parameters. Furthermore, a detailed study of the eclipses has the potential to test the boundary layer picture of X-ray emission from accreting WDs.


\begin{acknowledgements}
SM, GP, and KA acknowledge financial support from the European Research Council (ERC) under the European Union’s Horizon 2020 research and innovation program ``HotMilk'' (grant agreement No. 865637). GP acknowledges support from Bando per il Finanziamento della Ricerca Fondamentale 2022 dell’Istituto Nazionale di Astrofisica (INAF): GO Large program. MRM acknowledges support from NASA under grant GO1-22138X to UCLA. NR is supported by the ERC Consolidator Grant ``MAGNESIA'' under grant agreement No. 817661, and also partially supported by the program Unidad de Excelencia Maria de Maetzu de Maeztu CEX2020-001058-M. The work has made use of publicly available data from HEASARC Online Service, \emph{XMM-Newton} Science Analysis System (SAS) developed by the European Space Agency (ESA). {\it Software}: Python \citep{vanrossum2009}, Jupyter \citep{kluyver2016}, NumPy \citep{vanderwalt2011,harris2020}, matplotlib \citep{hunter2007}.

\end{acknowledgements}

%
%

\bibliographystyle{aa} 
\bibliography{refs}

\begin{thebibliography}{72}
\expandafter\ifx\csname natexlab\endcsname\relax\def\natexlab#1{#1}\fi

\bibitem[{{Aizu}(1973)}]{aizu1973}
{Aizu}, K. 1973,
  \href{http://dx.doi.org/10.1143/PTP.49.1184}{\color{magenta}Progress of
  Theoretical Physics},
  \href{https://ui.adsabs.harvard.edu/abs/1973PThPh..49.1184A}{49, 1184}

\bibitem[{{Arnaud}(1996)}]{arnaud1996}
{Arnaud}, K.~A. 1996, in Astronomical Society of the Pacific Conference Series,
  Vol. 101, Astronomical Data Analysis Software and Systems V, ed. G.~H.
  {Jacoby} \& J.~{Barnes},
  \href{https://ui.adsabs.harvard.edu/abs/1996ASPC..101...17A}{17}

\bibitem[{{Barlow} {et~al.}(2006){Barlow}, {Knigge}, {Bird}, {J Dean}, {Clark},
  {Hill}, {Molina}, \& {Sguera}}]{barlow2006}
{Barlow}, E.~J., {Knigge}, C., {Bird}, A.~J., {et~al.} 2006,
  \href{http://dx.doi.org/10.1111/j.1365-2966.2006.10836.x}{\color{magenta}\mnras},
  \href{https://ui.adsabs.harvard.edu/abs/2006MNRAS.372..224B}{372, 224}

\bibitem[{{Baumgartner} {et~al.}(2013){Baumgartner}, {Tueller}, {Markwardt},
  {Skinner}, {Barthelmy}, {Mushotzky}, {Evans}, \& {Gehrels}}]{baumgartner2013}
{Baumgartner}, W.~H., {Tueller}, J., {Markwardt}, C.~B., {et~al.} 2013,
  \href{http://dx.doi.org/10.1088/0067-0049/207/2/19}{\color{magenta}\apjs},
  \href{https://ui.adsabs.harvard.edu/abs/2013ApJS..207...19B}{207, 19}

\bibitem[{{Beuermann} {et~al.}(1999){Beuermann}, {Thomas}, {Reinsch},
  {Schwope}, {Tr{\"u}mper}, \& {Voges}}]{beuermann1999}
{Beuermann}, K., {Thomas}, H.~C., {Reinsch}, K., {et~al.} 1999, \aap,
  \href{https://ui.adsabs.harvard.edu/abs/1999A&A...347...47B}{347, 47}

\bibitem[{{Blondin} {et~al.}(1990){Blondin}, {Kallman}, {Fryxell}, \&
  {Taam}}]{blondin1990}
{Blondin}, J.~M., {Kallman}, T.~R., {Fryxell}, B.~A., \& {Taam}, R.~E. 1990,
  \href{http://dx.doi.org/10.1086/168865}{\color{magenta}\apj},
  \href{https://ui.adsabs.harvard.edu/abs/1990ApJ...356..591B}{356, 591}

\bibitem[{{Britt} {et~al.}(2014){Britt}, {Hynes}, {Johnson}, {Baldwin},
  {Jonker}, {Nelemans}, {Torres}, {Maccarone}, {Steeghs}, {Greiss}, {Heinke},
  {Bassa}, {Collazzi}, {Villar}, {Gabb}, \& {Gossen}}]{britt2014}
{Britt}, C.~T., {Hynes}, R.~I., {Johnson}, C.~B., {et~al.} 2014,
  \href{http://dx.doi.org/10.1088/0067-0049/214/1/10}{\color{magenta}\apjs},
  \href{https://ui.adsabs.harvard.edu/abs/2014ApJS..214...10B}{214, 10}

\bibitem[{{Coughenour} {et~al.}(2022){Coughenour}, {Tomsick}, {Shaw}, {Mukai},
  {Clavel}, {Hare}, {Krivonos}, \& {Fornasini}}]{coughenour2022}
{Coughenour}, B.~M., {Tomsick}, J.~A., {Shaw}, A.~W., {et~al.} 2022,
  \href{http://dx.doi.org/10.1093/mnras/stac263}{\color{magenta}\mnras},
  \href{https://ui.adsabs.harvard.edu/abs/2022MNRAS.511.4582C}{511, 4582}

\bibitem[{{Cropper}(1990)}]{copper1990}
{Cropper}, M. 1990,
  \href{http://dx.doi.org/10.1007/BF00177799}{\color{magenta}\ssr},
  \href{https://ui.adsabs.harvard.edu/abs/1990SSRv...54..195C}{54, 195}

\bibitem[{{D{\'\i}az Trigo} {et~al.}(2006){D{\'\i}az Trigo}, {Parmar},
  {Boirin}, {M{\'e}ndez}, \& {Kaastra}}]{diaz2006}
{D{\'\i}az Trigo}, M., {Parmar}, A.~N., {Boirin}, L., {M{\'e}ndez}, M., \&
  {Kaastra}, J.~S. 2006,
  \href{http://dx.doi.org/10.1051/0004-6361:20053586}{\color{magenta}\aap},
  \href{https://ui.adsabs.harvard.edu/abs/2006A&A...445..179D}{445, 179}

\bibitem[{{Edmonds} {et~al.}(2003{\natexlab{a}}){Edmonds}, {Gilliland},
  {Heinke}, \& {Grindlay}}]{edmonds2003b}
{Edmonds}, P.~D., {Gilliland}, R.~L., {Heinke}, C.~O., \& {Grindlay}, J.~E.
  2003{\natexlab{a}},
  \href{http://dx.doi.org/10.1086/378193}{\color{magenta}\apj},
  \href{https://ui.adsabs.harvard.edu/abs/2003ApJ...596.1177E}{596, 1177}

\bibitem[{{Edmonds} {et~al.}(2003{\natexlab{b}}){Edmonds}, {Gilliland},
  {Heinke}, \& {Grindlay}}]{edmonds2003a}
{Edmonds}, P.~D., {Gilliland}, R.~L., {Heinke}, C.~O., \& {Grindlay}, J.~E.
  2003{\natexlab{b}},
  \href{http://dx.doi.org/10.1086/378194}{\color{magenta}\apj},
  \href{https://ui.adsabs.harvard.edu/abs/2003ApJ...596.1197E}{596, 1197}

\bibitem[{{Evans} {et~al.}(2010){Evans}, {Primini}, {Glotfelty}, {Anderson},
  {Bonaventura}, {Chen}, {Davis}, {Doe}, {Evans}, {Fabbiano}, {Galle}, {Gibbs},
  {Grier}, {Hain}, {Hall}, {Harbo}, {He}, {Houck}, {Karovska}, {Kashyap},
  {Lauer}, {McCollough}, {McDowell}, {Miller}, {Mitschang}, {Morgan},
  {Mossman}, {Nichols}, {Nowak}, {Plummer}, {Refsdal}, {Rots}, {Siemiginowska},
  {Sundheim}, {Tibbetts}, {Van Stone}, {Winkelman}, \& {Zografou}}]{evans2010}
{Evans}, I.~N., {Primini}, F.~A., {Glotfelty}, K.~J., {et~al.} 2010,
  \href{http://dx.doi.org/10.1088/0067-0049/189/1/37}{\color{magenta}\apjs},
  \href{https://ui.adsabs.harvard.edu/abs/2010ApJS..189...37E}{189, 37}

\bibitem[{{Evans} {et~al.}(2020){Evans}, {Page}, {Osborne}, {Beardmore},
  {Willingale}, {Burrows}, {Kennea}, {Perri}, {Capalbi}, {Tagliaferri}, \&
  {Cenko}}]{evans2020}
{Evans}, P.~A., {Page}, K.~L., {Osborne}, J.~P., {et~al.} 2020,
  \href{http://dx.doi.org/10.3847/1538-4365/ab7db9}{\color{magenta}\apjs},
  \href{https://ui.adsabs.harvard.edu/abs/2020ApJS..247...54E}{247, 54}

\bibitem[{{Ezuka} \& {Ishida}(1999)}]{ezuka1999}
{Ezuka}, H. \& {Ishida}, M. 1999,
  \href{http://dx.doi.org/10.1086/313181}{\color{magenta}\apjs},
  \href{https://ui.adsabs.harvard.edu/abs/1999ApJS..120..277E}{120, 277}

\bibitem[{{Falanga} {et~al.}(2015){Falanga}, {Bozzo}, {Lutovinov},
  {Bonnet-Bidaud}, {Fetisova}, \& {Puls}}]{falanga2015}
{Falanga}, M., {Bozzo}, E., {Lutovinov}, A., {et~al.} 2015,
  \href{http://dx.doi.org/10.1051/0004-6361/201425191}{\color{magenta}\aap},
  \href{https://ui.adsabs.harvard.edu/abs/2015A&A...577A.130F}{577, A130}

\bibitem[{{Frank} {et~al.}(1987){Frank}, {King}, \& {Lasota}}]{frank1987}
{Frank}, J., {King}, A.~R., \& {Lasota}, J.~P. 1987, \aap,
  \href{https://ui.adsabs.harvard.edu/abs/1987A&A...178..137F}{178, 137}

\bibitem[{{Gaia Collaboration} {et~al.}(2016){Gaia Collaboration}, {Prusti},
  {de Bruijne}, {Brown}, {Vallenari}, {Babusiaux}, {Bailer-Jones}, {Bastian},
  {Biermann}, {Evans}, {Eyer}, {Jansen}, {Jordi}, {Klioner}, {Lammers},
  {Lindegren}, {Luri}, {Mignard}, {Milligan}, {Panem}, {Poinsignon},
  {Pourbaix}, {Randich}, {Sarri}, {Sartoretti}, {Siddiqui}, {Soubiran},
  {Valette}, {van Leeuwen}, {Walton}, {Aerts}, {Arenou}, {Cropper}, {Drimmel},
  {H{\o}g}, {Katz}, {Lattanzi}, {O'Mullane}, {Grebel}, {Holland}, {Huc},
  {Passot}, {Bramante}, {Cacciari}, {Casta{\~n}eda}, {Chaoul}, {Cheek}, {De
  Angeli}, {Fabricius}, {Guerra}, {Hern{\'a}ndez}, {Jean-Antoine-Piccolo},
  {Masana}, {Messineo}, {Mowlavi}, {Nienartowicz}, {Ord{\'o}{\~n}ez-Blanco},
  {Panuzzo}, {Portell}, {Richards}, {Riello}, {Seabroke}, {Tanga},
  {Th{\'e}venin}, {Torra}, {Els}, {Gracia-Abril}, {Comoretto},
  {Garcia-Reinaldos}, {Lock}, {Mercier}, {Altmann}, {Andrae}, {Astraatmadja},
  {Bellas-Velidis}, {Benson}, {Berthier}, {Blomme}, {Busso}, {Carry},
  {Cellino}, {Clementini}, {Cowell}, {Creevey}, {Cuypers}, {Davidson}, {De
  Ridder}, {de Torres}, {Delchambre}, {Dell'Oro}, {Ducourant}, {Fr{\'e}mat},
  {Garc{\'\i}a-Torres}, {Gosset}, {Halbwachs}, {Hambly}, {Harrison}, {Hauser},
  {Hestroffer}, {Hodgkin}, {Huckle}, {Hutton}, {Jasniewicz}, {Jordan},
  {Kontizas}, {Korn}, {Lanzafame}, {Manteiga}, {Moitinho}, {Muinonen},
  {Osinde}, {Pancino}, {Pauwels}, {Petit}, {Recio-Blanco}, {Robin}, {Sarro},
  {Siopis}, {Smith}, {Smith}, {Sozzetti}, {Thuillot}, {van Reeven}, {Viala},
  {Abbas}, {Abreu Aramburu}, {Accart}, {Aguado}, {Allan}, {Allasia},
  {Altavilla}, {{\'A}lvarez}, {Alves}, {Anderson}, {Andrei}, {Anglada Varela},
  {Antiche}, {Antoja}, {Ant{\'o}n}, {Arcay}, {Atzei}, {Ayache}, {Bach},
  {Baker}, {Balaguer-N{\'u}{\~n}ez}, {Barache}, {Barata}, {Barbier}, {Barblan},
  {Baroni}, {Barrado y Navascu{\'e}s}, {Barros}, {Barstow}, {Becciani},
  {Bellazzini}, {Bellei}, {Bello Garc{\'\i}a}, {Belokurov}, {Bendjoya},
  {Berihuete}, {Bianchi}, {Bienaym{\'e}}, {Billebaud}, {Blagorodnova},
  {Blanco-Cuaresma}, {Boch}, {Bombrun}, {Borrachero}, {Bouquillon}, {Bourda},
  {Bouy}, {Bragaglia}, {Breddels}, {Brouillet}, {Br{\"u}semeister},
  {Bucciarelli}, {Budnik}, {Burgess}, {Burgon}, {Burlacu}, {Busonero}, {Buzzi},
  {Caffau}, {Cambras}, {Campbell}, {Cancelliere}, {Cantat-Gaudin}, {Carlucci},
  {Carrasco}, {Castellani}, {Charlot}, {Charnas}, {Charvet}, {Chassat},
  {Chiavassa}, {Clotet}, {Cocozza}, {Collins}, {Collins}, {Costigan}, {Crifo},
  {Cross}, {Crosta}, {Crowley}, {Dafonte}, {Damerdji}, {Dapergolas}, {David},
  {David}, {De Cat}, {de Felice}, {de Laverny}, {De Luise}, {De March}, {de
  Martino}, {de Souza}, {Debosscher}, {del Pozo}, {Delbo}, {Delgado},
  {Delgado}, {di Marco}, {Di Matteo}, {Diakite}, {Distefano}, {Dolding}, {Dos
  Anjos}, {Drazinos}, {Dur{\'a}n}, {Dzigan}, {Ecale}, {Edvardsson}, {Enke},
  {Erdmann}, {Escolar}, {Espina}, {Evans}, {Eynard Bontemps}, {Fabre},
  {Fabrizio}, {Faigler}, {Falc{\~a}o}, {Farr{\`a}s Casas}, {Faye}, {Federici},
  {Fedorets}, {Fern{\'a}ndez-Hern{\'a}ndez}, {Fernique}, {Fienga}, {Figueras},
  {Filippi}, {Findeisen}, {Fonti}, {Fouesneau}, {Fraile}, {Fraser}, {Fuchs},
  {Furnell}, {Gai}, {Galleti}, {Galluccio}, {Garabato}, {Garc{\'\i}a-Sedano},
  {Gar{\'e}}, {Garofalo}, {Garralda}, {Gavras}, {Gerssen}, {Geyer}, {Gilmore},
  {Girona}, {Giuffrida}, {Gomes}, {Gonz{\'a}lez-Marcos},
  {Gonz{\'a}lez-N{\'u}{\~n}ez}, {Gonz{\'a}lez-Vidal}, {Granvik}, {Guerrier},
  {Guillout}, {Guiraud}, {G{\'u}rpide}, {Guti{\'e}rrez-S{\'a}nchez}, {Guy},
  {Haigron}, {Hatzidimitriou}, {Haywood}, {Heiter}, {Helmi}, {Hobbs},
  {Hofmann}, {Holl}, {Holland}, {Hunt}, {Hypki}, {Icardi}, {Irwin}, {Jevardat
  de Fombelle}, {Jofr{\'e}}, {Jonker}, {Jorissen}, {Julbe}, {Karampelas},
  {Kochoska}, {Kohley}, {Kolenberg}, {Kontizas}, {Koposov}, {Kordopatis},
  {Koubsky}, {Kowalczyk}, {Krone-Martins}, {Kudryashova}, {Kull}, {Bachchan},
  {Lacoste-Seris}, {Lanza}, {Lavigne}, {Le Poncin-Lafitte}, {Lebreton},
  {Lebzelter}, {Leccia}, {Leclerc}, {Lecoeur-Taibi}, {Lemaitre}, {Lenhardt},
  {Leroux}, {Liao}, {Licata}, {Lindstr{\o}m}, {Lister}, {Livanou}, {Lobel},
  {L{\"o}ffler}, {L{\'o}pez}, {Lopez-Lozano}, {Lorenz}, {Loureiro},
  {MacDonald}, {Magalh{\~a}es Fernandes}, {Managau}, {Mann}, {Mantelet},
  {Marchal}, {Marchant}, {Marconi}, {Marie}, {Marinoni}, {Marrese},
  {Marschalk{\'o}}, {Marshall}, {Mart{\'\i}n-Fleitas}, {Martino}, {Mary},
  {Matijevi{\v{c}}}, {Mazeh}, {McMillan}, {Messina}, {Mestre}, {Michalik},
  {Millar}, {Miranda}, {Molina}, {Molinaro}, {Molinaro}, {Moln{\'a}r},
  {Moniez}, {Montegriffo}, {Monteiro}, {Mor}, {Mora}, {Morbidelli}, {Morel},
  {Morgenthaler}, {Morley}, {Morris}, {Mulone}, {Muraveva}, {Musella},
  {Narbonne}, {Nelemans}, {Nicastro}, {Noval}, {Ord{\'e}novic},
  {Ordieres-Mer{\'e}}, {Osborne}, {Pagani}, {Pagano}, {Pailler}, {Palacin},
  {Palaversa}, {Parsons}, {Paulsen}, {Pecoraro}, {Pedrosa}, {Pentik{\"a}inen},
  {Pereira}, {Pichon}, {Piersimoni}, {Pineau}, {Plachy}, {Plum}, {Poujoulet},
  {Pr{\v{s}}a}, {Pulone}, {Ragaini}, {Rago}, {Rambaux}, {Ramos-Lerate},
  {Ranalli}, {Rauw}, {Read}, {Regibo}, {Renk}, {Reyl{\'e}}, {Ribeiro},
  {Rimoldini}, {Ripepi}, {Riva}, {Rixon}, {Roelens}, {Romero-G{\'o}mez},
  {Rowell}, {Royer}, {Rudolph}, {Ruiz-Dern}, {Sadowski}, {Sagrist{\`a}
  Sell{\'e}s}, {Sahlmann}, {Salgado}, {Salguero}, {Sarasso}, {Savietto},
  {Schnorhk}, {Schultheis}, {Sciacca}, {Segol}, {Segovia}, {Segransan},
  {Serpell}, {Shih}, {Smareglia}, {Smart}, {Smith}, {Solano}, {Solitro},
  {Sordo}, {Soria Nieto}, {Souchay}, {Spagna}, {Spoto}, {Stampa}, {Steele},
  {Steidelm{\"u}ller}, {Stephenson}, {Stoev}, {Suess}, {S{\"u}veges}, {Surdej},
  {Szabados}, {Szegedi-Elek}, {Tapiador}, {Taris}, {Tauran}, {Taylor},
  {Teixeira}, {Terrett}, {Tingley}, {Trager}, {Turon}, {Ulla}, {Utrilla},
  {Valentini}, {van Elteren}, {Van Hemelryck}, {van Leeuwen}, {Varadi},
  {Vecchiato}, {Veljanoski}, {Via}, {Vicente}, {Vogt}, {Voss}, {Votruba},
  {Voutsinas}, {Walmsley}, {Weiler}, {Weingrill}, {Werner}, {Wevers},
  {Whitehead}, {Wyrzykowski}, {Yoldas}, {{\v{Z}}erjal}, {Zucker}, {Zurbach},
  {Zwitter}, {Alecu}, {Allen}, {Allende Prieto}, {Amorim},
  {Anglada-Escud{\'e}}, {Arsenijevic}, {Azaz}, {Balm}, {Beck}, {Bernstein},
  {Bigot}, {Bijaoui}, {Blasco}, {Bonfigli}, {Bono}, {Boudreault}, {Bressan},
  {Brown}, {Brunet}, {Bunclark}, {Buonanno}, {Butkevich}, {Carret}, {Carrion},
  {Chemin}, {Ch{\'e}reau}, {Corcione}, {Darmigny}, {de Boer}, {de Teodoro}, {de
  Zeeuw}, {Delle Luche}, {Domingues}, {Dubath}, {Fodor}, {Fr{\'e}zouls},
  {Fries}, {Fustes}, {Fyfe}, {Gallardo}, {Gallegos}, {Gardiol}, {Gebran},
  {Gomboc}, {G{\'o}mez}, {Grux}, {Gueguen}, {Heyrovsky}, {Hoar}, {Iannicola},
  {Isasi Parache}, {Janotto}, {Joliet}, {Jonckheere}, {Keil}, {Kim},
  {Klagyivik}, {Klar}, {Knude}, {Kochukhov}, {Kolka}, {Kos}, {Kutka}, {Lainey},
  {LeBouquin}, {Liu}, {Loreggia}, {Makarov}, {Marseille}, {Martayan},
  {Martinez-Rubi}, {Massart}, {Meynadier}, {Mignot}, {Munari}, {Nguyen},
  {Nordlander}, {Ocvirk}, {O'Flaherty}, {Olias Sanz}, {Ortiz}, {Osorio},
  {Oszkiewicz}, {Ouzounis}, {Palmer}, {Park}, {Pasquato}, {Peltzer}, {Peralta},
  {P{\'e}turaud}, {Pieniluoma}, {Pigozzi}, {Poels}, {Prat}, {Prod'homme},
  {Raison}, {Rebordao}, {Risquez}, {Rocca-Volmerange}, {Rosen}, {Ruiz-Fuertes},
  {Russo}, {Sembay}, {Serraller Vizcaino}, {Short}, {Siebert}, {Silva},
  {Sinachopoulos}, {Slezak}, {Soffel}, {Sosnowska}, {Strai{\v{z}}ys}, {ter
  Linden}, {Terrell}, {Theil}, {Tiede}, {Troisi}, {Tsalmantza}, {Tur},
  {Vaccari}, {Vachier}, {Valles}, {Van Hamme}, {Veltz}, {Virtanen}, {Wallut},
  {Wichmann}, {Wilkinson}, {Ziaeepour}, \& {Zschocke}}]{gaia2016}
{Gaia Collaboration}, {Prusti}, T., {de Bruijne}, J.~H.~J., {et~al.} 2016,
  \href{http://dx.doi.org/10.1051/0004-6361/201629272}{\color{magenta}\aap},
  \href{https://ui.adsabs.harvard.edu/abs/2016A&A...595A...1G}{595, A1}

\bibitem[{{Gaia Collaboration} {et~al.}(2022){Gaia Collaboration}, {Vallenari},
  {Brown}, {Prusti}, {de Bruijne}, {Arenou}, {Babusiaux}, {Biermann},
  {Creevey}, {Ducourant}, {Evans}, {Eyer}, {Guerra}, {Hutton}, {Jordi},
  {Klioner}, {Lammers}, {Lindegren}, {Luri}, {Mignard}, {Panem}, {Pourbaix},
  {Randich}, {Sartoretti}, {Soubiran}, {Tanga}, {Walton}, {Bailer-Jones},
  {Bastian}, {Drimmel}, {Jansen}, {Katz}, {Lattanzi}, {van Leeuwen}, {Bakker},
  {Cacciari}, {Casta{\~n}eda}, {De Angeli}, {Fabricius}, {Fouesneau},
  {Fr{\'e}mat}, {Galluccio}, {Guerrier}, {Heiter}, {Masana}, {Messineo},
  {Mowlavi}, {Nicolas}, {Nienartowicz}, {Pailler}, {Panuzzo}, {Riclet}, {Roux},
  {Seabroke}, {Sordo{\o}rcit}, {Th{\'e}venin}, {Gracia-Abril}, {Portell},
  {Teyssier}, {Altmann}, {Andrae}, {Audard}, {Bellas-Velidis}, {Benson},
  {Berthier}, {Blomme}, {Burgess}, {Busonero}, {Busso}, {C{\'a}novas}, {Carry},
  {Cellino}, {Cheek}, {Clementini}, {Damerdji}, {Davidson}, {de Teodoro},
  {Nu{\~n}ez Campos}, {Delchambre}, {Dell'Oro}, {Esquej},
  {Fern{\'a}ndez-Hern{\'a}ndez}, {Fraile}, {Garabato}, {Garc{\'\i}a-Lario},
  {Gosset}, {Haigron}, {Halbwachs}, {Hambly}, {Harrison}, {Hern{\'a}ndez},
  {Hestroffer}, {Hodgkin}, {Holl}, {Jan{\ss}en}, {Jevardat de Fombelle},
  {Jordan}, {Krone-Martins}, {Lanzafame}, {L{\"o}ffler}, {Marchal}, {Marrese},
  {Moitinho}, {Muinonen}, {Osborne}, {Pancino}, {Pauwels}, {Recio-Blanco},
  {Reyl{\'e}}, {Riello}, {Rimoldini}, {Roegiers}, {Rybizki}, {Sarro}, {Siopis},
  {Smith}, {Sozzetti}, {Utrilla}, {van Leeuwen}, {Abbas}, {{\'A}brah{\'a}m},
  {Abreu Aramburu}, {Aerts}, {Aguado}, {Ajaj}, {Aldea-Montero}, {Altavilla},
  {{\'A}lvarez}, {Alves}, {Anders}, {Anderson}, {Anglada Varela}, {Antoja},
  {Baines}, {Baker}, {Balaguer-N{\'u}{\~n}ez}, {Balbinot}, {Balog}, {Barache},
  {Barbato}, {Barros}, {Barstow}, {Bartolom{\'e}}, {Bassilana}, {Bauchet},
  {Becciani}, {Bellazzini}, {Berihuete}, {Bernet}, {Bertone}, {Bianchi},
  {Binnenfeld}, {Blanco-Cuaresma}, {Blazere}, {Boch}, {Bombrun}, {Bossini},
  {Bouquillon}, {Bragaglia}, {Bramante}, {Breedt}, {Bressan}, {Brouillet},
  {Brugaletta}, {Bucciarelli}, {Burlacu}, {Butkevich}, {Buzzi}, {Caffau},
  {Cancelliere}, {Cantat-Gaudin}, {Carballo}, {Carlucci}, {Carnerero},
  {Carrasco}, {Casamiquela}, {Castellani}, {Castro-Ginard}, {Chaoul},
  {Charlot}, {Chemin}, {Chiaramida}, {Chiavassa}, {Chornay}, {Comoretto},
  {Contursi}, {Cooper}, {Cornez}, {Cowell}, {Crifo}, {Cropper}, {Crosta},
  {Crowley}, {Dafonte}, {Dapergolas}, {David}, {David}, {de Laverny}, {De
  Luise}, {De March}, {De Ridder}, {de Souza}, {de Torres}, {del Peloso}, {del
  Pozo}, {Delbo}, {Delgado}, {Delisle}, {Demouchy}, {Dharmawardena}, {Di
  Matteo}, {Diakite}, {Diener}, {Distefano}, {Dolding}, {Edvardsson}, {Enke},
  {Fabre}, {Fabrizio}, {Faigler}, {Fedorets}, {Fernique}, {Fienga}, {Figueras},
  {Fournier}, {Fouron}, {Fragkoudi}, {Gai}, {Garcia-Gutierrez},
  {Garcia-Reinaldos}, {Garc{\'\i}a-Torres}, {Garofalo}, {Gavel}, {Gavras},
  {Gerlach}, {Geyer}, {Giacobbe}, {Gilmore}, {Girona}, {Giuffrida}, {Gomel},
  {Gomez}, {Gonz{\'a}lez-N{\'u}{\~n}ez}, {Gonz{\'a}lez-Santamar{\'\i}a},
  {Gonz{\'a}lez-Vidal}, {Granvik}, {Guillout}, {Guiraud},
  {Guti{\'e}rrez-S{\'a}nchez}, {Guy}, {Hatzidimitriou}, {Hauser}, {Haywood},
  {Helmer}, {Helmi}, {Sarmiento}, {Hidalgo}, {Hilger}, {H{\l}adczuk}, {Hobbs},
  {Holland}, {Huckle}, {Jardine}, {Jasniewicz}, {Jean-Antoine Piccolo},
  {Jim{\'e}nez-Arranz}, {Jorissen}, {Juaristi Campillo}, {Julbe}, {Karbevska},
  {Kervella}, {Khanna}, {Kontizas}, {Kordopatis}, {Korn}, {K{\'o}sp{\'a}l},
  {Kostrzewa-Rutkowska}, {Kruszy{\'n}ska}, {Kun}, {Laizeau}, {Lambert},
  {Lanza}, {Lasne}, {Le Campion}, {Lebreton}, {Lebzelter}, {Leccia}, {Leclerc},
  {Lecoeur-Taibi}, {Liao}, {Licata}, {Lindstr{\o}m}, {Lister}, {Livanou},
  {Lobel}, {Lorca}, {Loup}, {Madrero Pardo}, {Magdaleno Romeo}, {Managau},
  {Mann}, {Manteiga}, {Marchant}, {Marconi}, {Marcos}, {Marcos Santos},
  {Mar{\'\i}n Pina}, {Marinoni}, {Marocco}, {Marshall}, {Polo},
  {Mart{\'\i}n-Fleitas}, {Marton}, {Mary}, {Masip}, {Massari},
  {Mastrobuono-Battisti}, {Mazeh}, {McMillan}, {Messina}, {Michalik}, {Millar},
  {Mints}, {Molina}, {Molinaro}, {Moln{\'a}r}, {Monari}, {Mongui{\'o}},
  {Montegriffo}, {Montero}, {Mor}, {Mora}, {Morbidelli}, {Morel}, {Morris},
  {Muraveva}, {Murphy}, {Musella}, {Nagy}, {Noval}, {Oca{\~n}a}, {Ogden},
  {Ordenovic}, {Osinde}, {Pagani}, {Pagano}, {Palaversa}, {Palicio},
  {Pallas-Quintela}, {Panahi}, {Payne-Wardenaar}, {Pe{\~n}alosa Esteller},
  {Penttil{\"a}}, {Pichon}, {Piersimoni}, {Pineau}, {Plachy}, {Plum}, {Poggio},
  {Pr{\v{s}}a}, {Pulone}, {Racero}, {Ragaini}, {Rainer}, {Raiteri}, {Rambaux},
  {Ramos}, {Ramos-Lerate}, {Re Fiorentin}, {Regibo}, {Richards}, {Rios Diaz},
  {Ripepi}, {Riva}, {Rix}, {Rixon}, {Robichon}, {Robin}, {Robin}, {Roelens},
  {Rogues}, {Rohrbasser}, {Romero-G{\'o}mez}, {Rowell}, {Royer}, {Ruz Mieres},
  {Rybicki}, {Sadowski}, {S{\'a}ez N{\'u}{\~n}ez}, {Sagrist{\`a} Sell{\'e}s},
  {Sahlmann}, {Salguero}, {Samaras}, {Sanchez Gimenez}, {Sanna},
  {Santove{\~n}a}, {Sarasso}, {Schultheis}, {Sciacca}, {Segol}, {Segovia},
  {S{\'e}gransan}, {Semeux}, {Shahaf}, {Siddiqui}, {Siebert}, {Siltala},
  {Silvelo}, {Slezak}, {Slezak}, {Smart}, {Snaith}, {Solano}, {Solitro},
  {Souami}, {Souchay}, {Spagna}, {Spina}, {Spoto}, {Steele},
  {Steidelm{\"u}ller}, {Stephenson}, {S{\"u}veges}, {Surdej}, {Szabados},
  {Szegedi-Elek}, {Taris}, {Taylo}, {Teixeira}, {Tolomei}, {Tonello}, {Torra},
  {Torra}, {Torralba Elipe}, {Trabucchi}, {Tsounis}, {Turon}, {Ulla}, {Unger},
  {Vaillant}, {van Dillen}, {van Reeven}, {Vanel}, {Vecchiato}, {Viala},
  {Vicente}, {Voutsinas}, {Weiler}, {Wevers}, {Wyrzykowski}, {Yoldas}, {Yvard},
  {Zhao}, {Zorec}, {Zucker}, \& {Zwitter}}]{gaia2022}
{Gaia Collaboration}, {Vallenari}, A., {Brown}, A.~G.~A., {et~al.} 2022,
  \href{https://ui.adsabs.harvard.edu/abs/2022arXiv220800211G}{arXiv e-prints,
  arXiv:2208.00211}

\bibitem[{{Gosnell} {et~al.}(2012){Gosnell}, {Pooley}, {Geller}, {Kalirai},
  {Mathieu}, {Frinchaboy}, \& {Ramirez-Ruiz}}]{gosnell2012}
{Gosnell}, N.~M., {Pooley}, D., {Geller}, A.~M., {et~al.} 2012,
  \href{http://dx.doi.org/10.1088/0004-637X/745/1/57}{\color{magenta}\apj},
  \href{https://ui.adsabs.harvard.edu/abs/2012ApJ...745...57G}{745, 57}

\bibitem[{{Grindlay} {et~al.}(2001{\natexlab{a}}){Grindlay}, {Heinke},
  {Edmonds}, \& {Murray}}]{grindlay2001a}
{Grindlay}, J.~E., {Heinke}, C., {Edmonds}, P.~D., \& {Murray}, S.~S.
  2001{\natexlab{a}},
  \href{http://dx.doi.org/10.1126/science.1061135}{\color{magenta}Science},
  \href{https://ui.adsabs.harvard.edu/abs/2001Sci...292.2290G}{292, 2290}

\bibitem[{{Grindlay} {et~al.}(2001{\natexlab{b}}){Grindlay}, {Heinke},
  {Edmonds}, {Murray}, \& {Cool}}]{grindlay2001b}
{Grindlay}, J.~E., {Heinke}, C.~O., {Edmonds}, P.~D., {Murray}, S.~S., \&
  {Cool}, A.~M. 2001{\natexlab{b}},
  \href{http://dx.doi.org/10.1086/338499}{\color{magenta}\apjl},
  \href{https://ui.adsabs.harvard.edu/abs/2001ApJ...563L..53G}{563, L53}

\bibitem[{{Haberl} \& {White}(1990)}]{haberl1990}
{Haberl}, F. \& {White}, N.~E. 1990,
  \href{http://dx.doi.org/10.1086/169187}{\color{magenta}\apj},
  \href{https://ui.adsabs.harvard.edu/abs/1990ApJ...361..225H}{361, 225}

\bibitem[{{Haberl} {et~al.}(1989){Haberl}, {White}, \& {Kallman}}]{haberl1989}
{Haberl}, F., {White}, N.~E., \& {Kallman}, T.~R. 1989,
  \href{http://dx.doi.org/10.1086/167714}{\color{magenta}\apj},
  \href{https://ui.adsabs.harvard.edu/abs/1989ApJ...343..409H}{343, 409}

\bibitem[{{Hailey} {et~al.}(2016){Hailey}, {Mori}, {Perez}, {Canipe}, {Hong},
  {Tomsick}, {Boggs}, {Christensen}, {Craig}, {Fornasini}, {Grindlay},
  {Harrison}, {Nynka}, {Rahoui}, {Stern}, {Zhang}, \& {Zhang}}]{hailey2016}
{Hailey}, C.~J., {Mori}, K., {Perez}, K., {et~al.} 2016,
  \href{http://dx.doi.org/10.3847/0004-637X/826/2/160}{\color{magenta}\apj},
  \href{https://ui.adsabs.harvard.edu/abs/2016ApJ...826..160H}{826, 160}

\bibitem[{Harris {et~al.}(2020)Harris, Millman, van~der Walt, Gommers,
  Virtanen, Cournapeau, Wieser, Taylor, Berg, Smith, Kern, Picus, Hoyer, van
  Kerkwijk, Brett, Haldane, del R{'{\i}}o, Wiebe, Peterson,
  G{'{e}}rard-Marchant, Sheppard, Reddy, Weckesser, Abbasi, Gohlke, \&
  Oliphant}]{harris2020}
Harris, C.~R., Millman, K.~J., van~der Walt, S.~J., {et~al.} 2020,
  \href{http://dx.doi.org/10.1038/s41586-020-2649-2}{\color{magenta}Nature},
  585, 357

\bibitem[{{Hellier}(1997)}]{hellier1997}
{Hellier}, C. 1997,
  \href{http://dx.doi.org/10.1093/mnras/291.1.71}{\color{magenta}\mnras},
  \href{https://ui.adsabs.harvard.edu/abs/1997MNRAS.291...71H}{291, 71}

\bibitem[{{Hong}(2012)}]{hong2012}
{Hong}, J. 2012,
  \href{http://dx.doi.org/10.1111/j.1365-2966.2012.22079.x}{\color{magenta}\mnras},
  \href{https://ui.adsabs.harvard.edu/abs/2012MNRAS.427.1633H}{427, 1633}

\bibitem[{{Hong} {et~al.}(2009){Hong}, {van den Berg}, {Grindlay}, \&
  {Laycock}}]{hong2009}
{Hong}, J.~S., {van den Berg}, M., {Grindlay}, J.~E., \& {Laycock}, S. 2009,
  \href{http://dx.doi.org/10.1088/0004-637X/706/1/223}{\color{magenta}\apj},
  \href{https://ui.adsabs.harvard.edu/abs/2009ApJ...706..223H}{706, 223}

\bibitem[{{Hunter}(2007)}]{hunter2007}
{Hunter}, J.~D. 2007,
  \href{http://dx.doi.org/10.1109/MCSE.2007.55}{\color{magenta}Computing in
  Science and Engineering},
  \href{https://ui.adsabs.harvard.edu/abs/2007CSE.....9...90H}{9, 90}

\bibitem[{{Jansen} {et~al.}(2001){Jansen}, {Lumb}, {Altieri}, {Clavel}, {Ehle},
  {Erd}, {Gabriel}, {Guainazzi}, {Gondoin}, {Much}, {Munoz}, {Santos},
  {Schartel}, {Texier}, \& {Vacanti}}]{jansen2001}
{Jansen}, F., {Lumb}, D., {Altieri}, B., {et~al.} 2001,
  \href{http://dx.doi.org/10.1051/0004-6361:20000036}{\color{magenta}\aap},
  \href{https://ui.adsabs.harvard.edu/abs/2001A&A...365L...1J}{365, L1}

\bibitem[{{Johnson} {et~al.}(2017){Johnson}, {Torres}, {Hynes}, {Jonker},
  {Heinke}, {Maccarone}, {Britt}, {Steeghs}, {Wevers}, \& {Wu}}]{johnson2017}
{Johnson}, C.~B., {Torres}, M.~A.~P., {Hynes}, R.~I., {et~al.} 2017,
  \href{http://dx.doi.org/10.1093/mnras/stw3063}{\color{magenta}\mnras},
  \href{https://ui.adsabs.harvard.edu/abs/2017MNRAS.466..129J}{466, 129}

\bibitem[{{Jonker} {et~al.}(2011){Jonker}, {Bassa}, {Nelemans}, {Steeghs},
  {Torres}, {Maccarone}, {Hynes}, {Greiss}, {Clem}, {Dieball}, {Mikles},
  {Britt}, {Gossen}, {Collazzi}, {Wijnands}, {In't Zand}, {M{\'e}ndez}, {Rea},
  {Kuulkers}, {Ratti}, {van Haaften}, {Heinke}, {{\"O}zel}, {Groot}, \&
  {Verbunt}}]{jonker2011}
{Jonker}, P.~G., {Bassa}, C.~G., {Nelemans}, G., {et~al.} 2011,
  \href{http://dx.doi.org/10.1088/0067-0049/194/2/18}{\color{magenta}\apjs},
  \href{https://ui.adsabs.harvard.edu/abs/2011ApJS..194...18J}{194, 18}

\bibitem[{{Jonker} {et~al.}(2014){Jonker}, {Torres}, {Hynes}, {Maccarone},
  {Steeghs}, {Greiss}, {Britt}, {Wu}, {Johnson}, {Nelemans}, \&
  {Heinke}}]{jonker2014}
{Jonker}, P.~G., {Torres}, M. A.~P., {Hynes}, R.~I., {et~al.} 2014,
  \href{http://dx.doi.org/10.1088/0067-0049/210/2/18}{\color{magenta}\apjs},
  \href{https://ui.adsabs.harvard.edu/abs/2014ApJS..210...18J}{210, 18}

\bibitem[{{Kepler} {et~al.}(2016){Kepler}, {Pelisoli}, {Koester}, {Ourique},
  {Romero}, {Reindl}, {Kleinman}, {Eisenstein}, {Valois}, \&
  {Amaral}}]{kepler2016}
{Kepler}, S.~O., {Pelisoli}, I., {Koester}, D., {et~al.} 2016,
  \href{http://dx.doi.org/10.1093/mnras/stv2526}{\color{magenta}\mnras},
  \href{https://ui.adsabs.harvard.edu/abs/2016MNRAS.455.3413K}{455, 3413}

\bibitem[{Kluyver {et~al.}(2016)Kluyver, Ragan-Kelley, P{\'e}rez, Granger,
  Bussonnier, Frederic, Kelley, Hamrick, Grout, Corlay, Ivanov, Avila, Abdalla,
  \& Willing}]{kluyver2016}
Kluyver, T., Ragan-Kelley, B., P{\'e}rez, F., {et~al.} 2016, in Positioning and
  Power in Academic Publishing: Players, Agents and Agendas, ed. F.~Loizides \&
  B.~Schmidt, IOS Press, \href{}{87 -- 90}

\bibitem[{{Krivonos} {et~al.}(2007){Krivonos}, {Revnivtsev}, {Churazov},
  {Sazonov}, {Grebenev}, \& {Sunyaev}}]{krivonos2007}
{Krivonos}, R., {Revnivtsev}, M., {Churazov}, E., {et~al.} 2007,
  \href{http://dx.doi.org/10.1051/0004-6361:20065626}{\color{magenta}\aap},
  \href{https://ui.adsabs.harvard.edu/abs/2007A&A...463..957K}{463, 957}

\bibitem[{{Manousakis} {et~al.}(2012){Manousakis}, {Walter}, \&
  {Blondin}}]{manousakis2012}
{Manousakis}, A., {Walter}, R., \& {Blondin}, J.~M. 2012,
  \href{http://dx.doi.org/10.1051/0004-6361/201219717}{\color{magenta}\aap},
  \href{https://ui.adsabs.harvard.edu/abs/2012A&A...547A..20M}{547, A20}

\bibitem[{{Meliani} {et~al.}(2000){Meliani}, {de Araujo}, \&
  {Aguiar}}]{melani2000}
{Meliani}, M.~T., {de Araujo}, J.~C.~N., \& {Aguiar}, O.~D. 2000, \aap,
  \href{https://ui.adsabs.harvard.edu/abs/2000A&A...358..417M}{358, 417}

\bibitem[{{Mondal} {et~al.}(2022){Mondal}, {Ponti}, {Haberl}, {Anastasopoulou},
  {Campana}, {Mori}, {Hailey}, \& {Rea}}]{mondal2022}
{Mondal}, S., {Ponti}, G., {Haberl}, F., {et~al.} 2022,
  \href{http://dx.doi.org/10.1051/0004-6361/202244264}{\color{magenta}\aap},
  \href{https://ui.adsabs.harvard.edu/abs/2022A&A...666A.150M}{666, A150}

\bibitem[{{Mukai}(2017)}]{mukai2017}
{Mukai}, K. 2017,
  \href{http://dx.doi.org/10.1088/1538-3873/aa6736}{\color{magenta}\pasp},
  \href{https://ui.adsabs.harvard.edu/abs/2017PASP..129f2001M}{129, 062001}

\bibitem[{{Mukai} {et~al.}(2009){Mukai}, {Zietsman}, \& {Still}}]{mukai2009}
{Mukai}, K., {Zietsman}, E., \& {Still}, M. 2009,
  \href{http://dx.doi.org/10.1088/0004-637X/707/1/652}{\color{magenta}\apj},
  \href{https://ui.adsabs.harvard.edu/abs/2009ApJ...707..652M}{707, 652}

\bibitem[{{Muno} {et~al.}(2003){Muno}, {Baganoff}, {Bautz}, {Brandt}, {Broos},
  {Feigelson}, {Garmire}, {Morris}, {Ricker}, \& {Townsley}}]{muno2003}
{Muno}, M.~P., {Baganoff}, F.~K., {Bautz}, M.~W., {et~al.} 2003,
  \href{http://dx.doi.org/10.1086/374639}{\color{magenta}\apj},
  \href{https://ui.adsabs.harvard.edu/abs/2003ApJ...589..225M}{589, 225}

\bibitem[{{Pandel} {et~al.}(2002){Pandel}, {Cordova}, {Shirey}, {Ramsay},
  {Cropper}, {Mason}, {Much}, \& {Kilkenny}}]{pandel2002}
{Pandel}, D., {Cordova}, F.~A., {Shirey}, R.~E., {et~al.} 2002,
  \href{http://dx.doi.org/10.1046/j.1365-8711.2002.05279.x}{\color{magenta}\mnras},
  \href{https://ui.adsabs.harvard.edu/abs/2002MNRAS.332..116P}{332, 116}

\bibitem[{{Patterson}(1994)}]{petterson1994}
{Patterson}, J. 1994,
  \href{http://dx.doi.org/10.1086/133375}{\color{magenta}\pasp},
  \href{https://ui.adsabs.harvard.edu/abs/1994PASP..106..209P}{106, 209}

\bibitem[{{Perez} {et~al.}(2015){Perez}, {Hailey}, {Bauer}, {Krivonos}, {Mori},
  {Baganoff}, {Barri{\`e}re}, {Boggs}, {Christensen}, {Craig}, {Grefenstette},
  {Grindlay}, {Harrison}, {Hong}, {Madsen}, {Nynka}, {Stern}, {Tomsick}, {Wik},
  {Zhang}, {Zhang}, \& {Zoglauer}}]{perez2015}
{Perez}, K., {Hailey}, C.~J., {Bauer}, F.~E., {et~al.} 2015,
  \href{http://dx.doi.org/10.1038/nature14353}{\color{magenta}\nat},
  \href{https://ui.adsabs.harvard.edu/abs/2015Natur.520..646P}{520, 646}

\bibitem[{{Ponti} {et~al.}(2016){Ponti}, {Bianchi}, {Mu{\~n}oz-Darias}, {De},
  {Fender}, \& {Merloni}}]{ponti2016}
{Ponti}, G., {Bianchi}, S., {Mu{\~n}oz-Darias}, T., {et~al.} 2016,
  \href{http://dx.doi.org/10.1002/asna.201612339}{\color{magenta}Astronomische
  Nachrichten},
  \href{https://ui.adsabs.harvard.edu/abs/2016AN....337..512P}{337, 512}

\bibitem[{{Ponti} {et~al.}(2019){Ponti}, {Hofmann}, {Churazov}, {Morris},
  {Haberl}, {Nandra}, {Terrier}, {Clavel}, \& {Goldwurm}}]{ponti2019}
{Ponti}, G., {Hofmann}, F., {Churazov}, E., {et~al.} 2019,
  \href{http://dx.doi.org/10.1038/s41586-019-1009-6}{\color{magenta}\nat},
  \href{https://ui.adsabs.harvard.edu/abs/2019Natur.567..347P}{567, 347}

\bibitem[{{Ponti} {et~al.}(2013){Ponti}, {Morris}, {Terrier}, \&
  {Goldwurm}}]{ponti2013}
{Ponti}, G., {Morris}, M.~R., {Terrier}, R., \& {Goldwurm}, A. 2013, in
  Astrophysics and Space Science Proceedings, Vol.~34, Cosmic Rays in
  Star-Forming Environments, ed. D.~F. {Torres} \& O.~{Reimer},
  \href{https://ui.adsabs.harvard.edu/abs/2013ASSP...34..331P}{331}

\bibitem[{{Ponti} {et~al.}(2015){Ponti}, {Morris}, {Terrier}, {Haberl},
  {Sturm}, {Clavel}, {Soldi}, {Goldwurm}, {Predehl}, {Nandra}, {B{\'e}langer},
  {Warwick}, \& {Tatischeff}}]{ponti2015}
{Ponti}, G., {Morris}, M.~R., {Terrier}, R., {et~al.} 2015,
  \href{http://dx.doi.org/10.1093/mnras/stv1331}{\color{magenta}\mnras},
  \href{https://ui.adsabs.harvard.edu/abs/2015MNRAS.453..172P}{453, 172}

\bibitem[{{Pretorius} {et~al.}(2007){Pretorius}, {Knigge}, {O'Donoghue},
  {Henry}, {Gioia}, \& {Mullis}}]{pretorius2007}
{Pretorius}, M.~L., {Knigge}, C., {O'Donoghue}, D., {et~al.} 2007,
  \href{http://dx.doi.org/10.1111/j.1365-2966.2007.12461.x}{\color{magenta}\mnras},
  \href{https://ui.adsabs.harvard.edu/abs/2007MNRAS.382.1279P}{382, 1279}

\bibitem[{{Ramsay}(2000)}]{ramsay2000}
{Ramsay}, G. 2000,
  \href{http://dx.doi.org/10.1046/j.1365-8711.2000.03239.x}{\color{magenta}\mnras},
  \href{https://ui.adsabs.harvard.edu/abs/2000MNRAS.314..403R}{314, 403}

\bibitem[{{Ramsay} \& {Cropper}(2007)}]{ramsay2007}
{Ramsay}, G. \& {Cropper}, M. 2007,
  \href{http://dx.doi.org/10.1111/j.1365-2966.2007.12011.x}{\color{magenta}\mnras},
  \href{https://ui.adsabs.harvard.edu/abs/2007MNRAS.379.1209R}{379, 1209}

\bibitem[{{Revnivtsev} {et~al.}(2009){Revnivtsev}, {Sazonov}, {Churazov},
  {Forman}, {Vikhlinin}, \& {Sunyaev}}]{revnivtsev2009}
{Revnivtsev}, M., {Sazonov}, S., {Churazov}, E., {et~al.} 2009,
  \href{http://dx.doi.org/10.1038/nature07946}{\color{magenta}\nat},
  \href{https://ui.adsabs.harvard.edu/abs/2009Natur.458.1142R}{458, 1142}

\bibitem[{{Ritter} \& {Kolb}(2003)}]{ritter2003}
{Ritter}, H. \& {Kolb}, U. 2003,
  \href{http://dx.doi.org/10.1051/0004-6361:20030330}{\color{magenta}\aap},
  \href{https://ui.adsabs.harvard.edu/abs/2003A&A...404..301R}{404, 301}

\bibitem[{{Saxton} {et~al.}(2005){Saxton}, {Wu}, {Cropper}, \&
  {Ramsay}}]{saxton2005}
{Saxton}, C.~J., {Wu}, K., {Cropper}, M., \& {Ramsay}, G. 2005,
  \href{http://dx.doi.org/10.1111/j.1365-2966.2005.09103.x}{\color{magenta}\mnras},
  \href{https://ui.adsabs.harvard.edu/abs/2005MNRAS.360.1091S}{360, 1091}

\bibitem[{{Schwope} {et~al.}(2001){Schwope}, {Schwarz}, {Sirk}, \&
  {Howell}}]{schwope2001}
{Schwope}, A.~D., {Schwarz}, R., {Sirk}, M., \& {Howell}, S.~B. 2001,
  \href{http://dx.doi.org/10.1051/0004-6361:20010838}{\color{magenta}\aap},
  \href{https://ui.adsabs.harvard.edu/abs/2001A&A...375..419S}{375, 419}

\bibitem[{{Shaw} {et~al.}(2020){Shaw}, {Heinke}, {Mukai}, {Tomsick},
  {Doroshenko}, {Suleimanov}, {Buisson}, {Gandhi}, {Grefenstette}, {Hare},
  {Jiang}, {Ludlam}, {Rana}, \& {Sivakoff}}]{shaw2020}
{Shaw}, A.~W., {Heinke}, C.~O., {Mukai}, K., {et~al.} 2020,
  \href{http://dx.doi.org/10.1093/mnras/staa2592}{\color{magenta}\mnras},
  \href{https://ui.adsabs.harvard.edu/abs/2020MNRAS.498.3457S}{498, 3457}

\bibitem[{{Suleimanov} {et~al.}(2016){Suleimanov}, {Doroshenko}, {Ducci},
  {Zhukov}, \& {Werner}}]{suleimanov2016}
{Suleimanov}, V., {Doroshenko}, V., {Ducci}, L., {Zhukov}, G.~V., \& {Werner},
  K. 2016,
  \href{http://dx.doi.org/10.1051/0004-6361/201628301}{\color{magenta}\aap},
  \href{https://ui.adsabs.harvard.edu/abs/2016A&A...591A..35S}{591, A35}

\bibitem[{{Suleimanov} {et~al.}(2019){Suleimanov}, {Doroshenko}, \&
  {Werner}}]{suleimanov2019}
{Suleimanov}, V.~F., {Doroshenko}, V., \& {Werner}, K. 2019,
  \href{http://dx.doi.org/10.1093/mnras/sty2952}{\color{magenta}\mnras},
  \href{https://ui.adsabs.harvard.edu/abs/2019MNRAS.482.3622S}{482, 3622}

\bibitem[{{Udalski} {et~al.}(2012){Udalski}, {Kowalczyk}, {Soszy{\'n}ski},
  {Poleski}, {Szyma{\'n}ski}, {Kubiak}, {Pietrzy{\'n}ski}, {Koz{\l}owski},
  {Pietrukowicz}, {Ulaczyk}, {Skowron}, \& {Wyrzykowski}}]{udalski2012}
{Udalski}, A., {Kowalczyk}, K., {Soszy{\'n}ski}, I., {et~al.} 2012, \actaa,
  \href{https://ui.adsabs.harvard.edu/abs/2012AcA....62..133U}{62, 133}

\bibitem[{{van der Walt} {et~al.}(2011){van der Walt}, {Colbert}, \&
  {Varoquaux}}]{vanderwalt2011}
{van der Walt}, S., {Colbert}, S.~C., \& {Varoquaux}, G. 2011,
  \href{http://dx.doi.org/10.1109/MCSE.2011.37}{\color{magenta}Computing in
  Science and Engineering},
  \href{https://ui.adsabs.harvard.edu/abs/2011CSE....13b..22V}{13, 22}

\bibitem[{Van~Rossum \& Drake(2009)}]{vanrossum2009}
Van~Rossum, G. \& Drake, F.~L. 2009, Python 3 Reference Manual (Scotts Valley,
  CA: CreateSpace)

\bibitem[{{Warwick} {et~al.}(1985){Warwick}, {Turner}, {Watson}, \&
  {Willingale}}]{warwick1985}
{Warwick}, R.~S., {Turner}, M.~J.~L., {Watson}, M.~G., \& {Willingale}, R.
  1985, \href{http://dx.doi.org/10.1038/317218a0}{\color{magenta}\nat},
  \href{https://ui.adsabs.harvard.edu/abs/1985Natur.317..218W}{317, 218}

\bibitem[{{Webb} {et~al.}(2020){Webb}, {Coriat}, {Traulsen}, {Ballet}, {Motch},
  {Carrera}, {Koliopanos}, {Authier}, {de la Calle}, {Ceballos}, {Colomo},
  {Chuard}, {Freyberg}, {Garcia}, {Kolehmainen}, {Lamer}, {Lin}, {Maggi},
  {Michel}, {Page}, {Page}, {Perea-Calderon}, {Pineau}, {Rodriguez}, {Rosen},
  {Santos Lleo}, {Saxton}, {Schwope}, {Tom{\'a}s}, {Watson}, \&
  {Zakardjian}}]{webb2020}
{Webb}, N.~A., {Coriat}, M., {Traulsen}, I., {et~al.} 2020,
  \href{http://dx.doi.org/10.1051/0004-6361/201937353}{\color{magenta}\aap},
  \href{https://ui.adsabs.harvard.edu/abs/2020A&A...641A.136W}{641, A136}

\bibitem[{{White} \& {Mason}(1985)}]{white1985}
{White}, N.~E. \& {Mason}, K.~O. 1985,
  \href{http://dx.doi.org/10.1007/BF00212883}{\color{magenta}\ssr},
  \href{https://ui.adsabs.harvard.edu/abs/1985SSRv...40..167W}{40, 167}

\bibitem[{{Willingale} {et~al.}(2013){Willingale}, {Starling}, {Beardmore},
  {Tanvir}, \& {O'Brien}}]{willingale2013}
{Willingale}, R., {Starling}, R.~L.~C., {Beardmore}, A.~P., {Tanvir}, N.~R., \&
  {O'Brien}, P.~T. 2013,
  \href{http://dx.doi.org/10.1093/mnras/stt175}{\color{magenta}\mnras},
  \href{https://ui.adsabs.harvard.edu/abs/2013MNRAS.431..394W}{431, 394}

\bibitem[{{Wilms} {et~al.}(2000){Wilms}, {Allen}, \& {McCray}}]{wilms2000}
{Wilms}, J., {Allen}, A., \& {McCray}, R. 2000,
  \href{http://dx.doi.org/10.1086/317016}{\color{magenta}\apj},
  \href{https://ui.adsabs.harvard.edu/abs/2000ApJ...542..914W}{542, 914}

\bibitem[{{Yamauchi} {et~al.}(2016){Yamauchi}, {Nobukawa}, {Nobukawa},
  {Uchiyama}, \& {Koyama}}]{yamauchi2016}
{Yamauchi}, S., {Nobukawa}, K.~K., {Nobukawa}, M., {Uchiyama}, H., \& {Koyama},
  K. 2016, \href{http://dx.doi.org/10.1093/pasj/psw057}{\color{magenta}\pasj},
  \href{https://ui.adsabs.harvard.edu/abs/2016PASJ...68...59Y}{68, 59}

\bibitem[{{Yuasa} {et~al.}(2012){Yuasa}, {Makishima}, \&
  {Nakazawa}}]{yuasa2012}
{Yuasa}, T., {Makishima}, K., \& {Nakazawa}, K. 2012,
  \href{http://dx.doi.org/10.1088/0004-637X/753/2/129}{\color{magenta}\apj},
  \href{https://ui.adsabs.harvard.edu/abs/2012ApJ...753..129Y}{753, 129}

\bibitem[{{Zorotovic} {et~al.}(2011){Zorotovic}, {Schreiber}, \&
  {G{\"a}nsicke}}]{zorotovic2011}
{Zorotovic}, M., {Schreiber}, M.~R., \& {G{\"a}nsicke}, B.~T. 2011,
  \href{http://dx.doi.org/10.1051/0004-6361/201116626}{\color{magenta}\aap},
  \href{https://ui.adsabs.harvard.edu/abs/2011A&A...536A..42Z}{536, A42}

\bibitem[{{Zou} {et~al.}(2020){Zou}, {Zhou}, \& {Huang}}]{zuo2020}
{Zou}, Z.-C., {Zhou}, X.-L., \& {Huang}, Y.-F. 2020,
  \href{http://dx.doi.org/10.1088/1674-4527/20/9/137}{\color{magenta}Research
  in Astronomy and Astrophysics},
  \href{https://ui.adsabs.harvard.edu/abs/2020RAA....20..137Z}{20, 137}

\end{thebibliography}

\end{document}